\documentclass[aps,prd,showpacs,twocolumn,showkeys,preprintnumbers]{revtex4}
\usepackage{graphicx}                                 
\usepackage{amsmath,amsfonts}
\graphicspath{{./Figures/}}                           

\usepackage{epsfig}
\usepackage{amssymb}
\usepackage{amsfonts}
\usepackage{float}
\newcommand{\half}{\frac{1}{2}}

\newcommand{\Tr}{\mbox{Tr}}

\newcommand{\bc}{\texttt{\,bc\,}}                         
\newcommand{\fc}{\texttt{\,fc\,}}                         
\newcommand{\noi}{\noindent}
\newcommand{\beq}{\begin{equation}}
\newcommand{\eeq}{\end{equation}}
\newcommand{\bea}{\begin{eqnarray}}
\newcommand{\eea}{\end{eqnarray}}
\newcommand{\beas}{\begin{eqnarray*}}
\newcommand{\eeas}{\end{eqnarray*}}
\newcommand{\aleq}{\mbox{}_{\textstyle \sim}^{\textstyle < }}

\newcommand{\eq}{\begin{equation}}
\newcommand{\en}{\end{equation}}
\newcommand{\eqa}{\begin{eqnarray}}
\newcommand{\ena}{\end{eqnarray}}
\newcommand{\Fig}[1]{Fig.~\ref{#1}}
\newcommand{\Tab}[1]{Table~\ref{#1}}
\newcommand{\Sec}[1]{Sec.~\ref{#1}}
\newcommand{\Eq}[1]{Eq.~(\ref{#1})}

\begin{document}

\preprint{HU-EP-05/69, 2nd revised version}

\title{Landau gauge ghost and gluon propagators in
$SU(2)$ lattice gauge theory: Gribov ambiguity revisited}
\author{I.L.~Bogolubsky}
\affiliation{Joint Institute for Nuclear Research, 141980 Dubna,
Russia}
\author{G.~Burgio}
\affiliation{Humboldt-Universit\"at zu Berlin, Institut f\"ur Physik,
12489 Berlin, Germany}
\author{V.K.~Mitrjushkin}
\affiliation{Joint Institute for Nuclear Research,
141980 Dubna, Russia}
\affiliation{Institute of Theoretical and Experimental Physics,
Moscow, Russia}
\author{M. M\"uller--Preussker}
\affiliation{Humboldt-Universit\"at zu Berlin, Institut f\"ur Physik,
12489 Berlin, Germany}

\date{May 27, 2006}

\begin{abstract}

We reinvestigate the problem of Gribov ambiguities within the Landau
(or Lorentz) gauge for the ghost and gluon propagators in pure
$SU(2)$ lattice gauge theory.  We make use of the full symmetry
group of the action taking into account {\it large}, i.e. non-periodic
$\mathbb{Z}(2)$ gauge transformations leaving lattice
plaquettes invariant. Enlarging in this way the gauge
orbits for any given gauge field configuration the Landau gauge can be fixed
at higher local extrema of the gauge functional in comparison with
standard (overrelaxation) techniques. This has a clearly visible effect
not only for the ghost propagator at small momenta but also for the gluon
propagator, 
in contrast to the common belief.

\end{abstract}

\keywords{Landau gauge, Gribov problem, gluon propagator, ghost propagator}
\pacs{11.15.Ha, 12.38.Gc, 12.38.Aw}
\maketitle

\section{Introduction}
\label{sec:introduction}
It is an important task to compute (Landau or Coulomb) gauge
gluon, ghost, fermion propagators and the basic vertex functions
from non-perturbative approaches to $SU(N)$ gauge theories, like
Dyson-Schwinger equations or the lattice formulation. On one hand one
is interested in their behavior in the infrared limit in order to
extract non-perturbative informations on various observables, e.g.
the QCD running couplings $\alpha_s(q^2)$, to
understand quark and gluon confinement within the
Gribov-Zwanziger scenario
\cite{Gribov:1977wm,Zwanziger:1993dh,Zwanziger:2003cf}, or to check
the Kugo-Ojima confinement criterion for the absence of colored
states \cite{Kugo:1979gm}. On the other hand it is
technically important to see to what extent these different
non-perturbative approaches provide results consistent with
each other in the non-perturbative region, i.e. at low momenta. At
present we are still far from drawing final conclusions in this
respect.  In particular the Dyson-Schwinger approach
\cite{Alkofer:2000wg,Fischer:2003rp,Lerche:2002ep}, always relying
on a truncated set of equations, provides results which look quite
different in the infinite volume limit compared with those
obtained on a torus
\cite{Fischer:2002eq,Fischer:2002hn,Fischer:2005ui}, while the latter
show at least qualitative agreement with recent results of numerical
lattice simulations \cite{Sternbeck:2005tk}.

It is well known that gauge fixing in the non-perturbative range is faced
with the Gribov ambiguity problem, which means that there can be many
gauge copies for a given gauge field satisfying the Landau gauge condition
$~\partial_{\mu} A_{\mu} =0~$ within the Gribov region, the latter defined by
the positivity of the Landau gauge Faddeev-Popov operator. In recent
years one has checked in greater detail how strong Gribov copies can influence
the infrared behavior especially of the gluon and ghost propagators.
Several groups of authors came to the conclusion that while there is a
clearly visible influence on the ghost propagator, the gluon propagator
seems only weakly affected 
\cite{Cucchieri:1997dx,Bakeev:2003rr,Nakajima:2003my,%
Silva:2004bv,Sternbeck:2005tk}.

Recently Zwanziger has argued that in the infinite volume limit the influence
of Gribov copies: ``... might be negligible, i.e. all averages taken over the Gribov region
should become equal to averages over the fundamental modular region''~\cite{Zwanziger:2003cf}. 
However, in practical lattice
simulations we are always restricted to finite volumes. Thus, Gribov copies
have to be taken into account properly before extrapolating to the infrared
and infinite volume.

In this paper we present a reinvestigation of the Gribov copy problem
for the $SU(2)$ case. The usual way to fix the (Landau) gauge on the
lattice is to simulate the path integral in its gauge invariant form.
Subsequently each of the produced lattice gauge fields
$~U \equiv \{U_{x,\mu}\}~$ is subjected to an iterative procedure maximizing
the gauge functional
\bea  \nonumber
F(g) &=& \frac{1}{d V} \sum_{x,\mu} \frac{1}{2} \mbox{Tr} ~U^g_{x,\mu}\,, \\
U^g_{x,\mu} &=& g(x) ~U_{x,\mu} ~g^{\dagger}(x+{\hat\mu})
\label{eq:functional}
\eea
with respect to local gauge transformations $~g \equiv \{g(x) \in SU(2)\}$.
$~V = L^{d}~$ denotes the number of lattice sites in $~d=4~$ dimensions.
The local maxima of $~F(g)~$ satisfy the differential lattice Landau gauge
transversality condition
\beq
(\partial_{\mu} A^g_{\mu})(x) =
       A^g_{\mu}(x+{\hat \mu}/2)-A^g_{\mu}(x-{\hat\mu}/2) = 0\,,
\label{eq:transversality}
\eeq
where the lattice gauge potentials are
\beq
A_{\mu}(x+ {\hat \mu}/2) =
       \frac{1}{2i}(U_{x,\mu}-U_{x,\mu}^{\dagger})\,.
\label{eq:potential}
\eeq
The standard procedure assumes {\it periodic gauge transformations}
and employs the {\it overrelaxation algorithm}. In what follows we shall
abbreviate it by {\it SOR}. The influence of Gribov copies
can be easily studied by taking various initial random gauge copies of the
gauge field configurations before subjecting them to the SOR algorithm.

At this place it is worth to note that the widely - at least till now -
accepted approach to compute e.g. a gauge-variant propagator $G$ is to choose
always the gauge copy with the highest value of local maxima $F_{max}$
(or the {\it best copy}) found for the gauge functional (\ref{eq:functional}).
One can then hope to have found a copy belonging to the so-called
{\it fundamental modular region} or at least being not far from it.
In order to find the {\it best copy} for each
thermalized gauge field configuration one needs to compare
$F_{max}$-values for a pretty large amount of gauge copies, which
is a rather time consuming procedure. A reasonable question is if
the use of only one gauge copy (the {\it first copy}) provides us
with the same - within errorbars - values of the propagator as the
use of the best copy. This logic brings us to compare the
propagator calculated on best copies ($G^{(bc)}$) with that on the
first copies ($G^{(fc)}$). The relative deviation $~\delta G
\equiv |(G^{(fc)}-G^{(bc)})/G^{(bc)}|~$ provides then a useful
quantitive measure of the Gribov ambiguity of the quantity under
consideration. We shall discuss this measure throughout the
present paper.

Of course, one can enhance the effect of Gribov copies by comparing instead
the best copies with the {\it worst copies}, i.e. with those having the
smallest values $F_{max}$ found from the repeated use of a given maximization
method. This attitude has been taken in Ref. \cite{Silva:2004bv} in order
to highlight a Gribov copy effect for the gluon propagator.

In Refs. \cite{Bakeev:2003rr} for $SU(2)$ and \cite{Sternbeck:2005tk} for
$SU(3)$ some of us already have thoroughly discussed the impact of Gribov
copies  within the SOR framework by comparing first and best copies.
From this point of view the gluon propagator did not
depend on the copies within the statistical noise,
whereas the ghost propagator clearly was depending on them in the
infrared. But the data for the ghost propagator obtained for different
lattice sizes showed an indication for a weakening of the dependence on
the choice of Gribov copies for increasing lattice size at fixed momentum,
in agreement with Zwanziger's claim~\cite{Zwanziger:2003cf}.

Here we enlarge the class of possible gauge transformations by taking into
account also {\it non-periodic} center gauge transformations. This will
allow us to maximize further the gauge functional and to see a quite strong
Gribov copy effect also for the gluon propagator at finite (lattice) volumes.

In \Sec{sec:gaugefixing} we shall explain the improved gauge fixing
procedure. In \Sec{sec:propagators} we define the propagators to be
calculated. In \Sec{sec:results} we are going to present
our results for the gluon and ghost propagators, whereas in
\Sec{sec:conclusions} the conclusions will be drawn.

\section{Improved gauge fixing}
\label{sec:gaugefixing}

We shall deal all the time with $SU(2)$ pure gauge lattice fields in four
Euclidean dimensions produced by means of Monte Carlo simulations with
the standard Wilson plaquette action. We restrict ourselves to the
confinement phase at $T=0$.

To fix the gauge we employ the standard Los Alamos type overrelaxation
with $\omega = 1.7$.

Our generalization of the standard gauge fixing procedure SOR
comes from the simple observation that gauge covariance for
periodic $SU(2)$ gauge fields on a $d$-dimensional torus of
extension $L^d$ allows gauge transformations which are not
necessarily periodic but can differ by a group center element at
the boundary: \beq g(x+L\hat{\nu}) = z_{\nu} g(x)\,, z_{\nu}=\pm 1
\in \mathbb{Z}(2)\,. \eeq In light of this it is legitimate to
allow, during the maximization of the gauge functional in the
gauge fixing procedure, for gauge tranformations which differ by a
sign when winding around a boundary. Let $\nu$ be the direction of
such boundary. Any such gauge transformation can be decomposed
into a standard periodic gauge transformation (which we may call a
``small'' one) and a flip of all links $~U_{\nu}(x) \to -
~U_{\nu}(x)~$ of a 3-plane at a given fixed $~x_{\nu}$. Given a
``small'' random gauge copy of the configuration we have thus
performed a pre-conditioning step for the gauge functional by
sweeping in every direction all 3-planes in succession and
comparing the value of the flipped with the unflipped gauge
functional. The flip is accepted if the gauge functional
increases. It is easy to see that such a procedure is independent
of the order of choosing the 3-planes and that only one sweep
through the lattice is required to maximize the functional. The gauge copy
obtained at the end of this procedure is then used as a starting
point for a standard maximization procedure. We call the whole
procedure FOR.

Analogously to the SOR method the FOR procedure can be repeated with different
initial random gauges in order to find a best copy (\bc{}) in comparison e.g.
with the first random copy (\fc{}). We shall check the convergence of the
\bc{}-propagator results for the best copies as a function of the number
$~n_{copy}~$ of random  initial copies.

\section{Gluon and ghost propagators}
\label{sec:propagators}

We turn now to the computation of the gauge variant gluon and
ghost propagators within the Landau gauge.

The lattice gluon propagator $~D^{ab}_{\mu\nu}(p)~$ is taken as the
Fourier transform of the gluon two-point function, {\it i.e.} the
expectation value
\bea
D^{ab}_{\mu\nu}(p) &=&
\left\langle \widetilde{A}^a_{\mu}(\hat{k}) \widetilde{A}^b_{\nu}(-\hat{k})
\right\rangle_U
\\   &=& \delta^{ab} \left(\delta_{\mu\nu} - \frac{p_{\mu}~p_{\nu}}{p^2}
                           \right) D(p) \,. \nonumber
\label{eq:D-def}
\eea
$\widetilde{A}^a_{\mu}(\hat{k})~$ is the Fourier transform of the lattice gauge
potential $~A^a_\mu(x+\hat{\mu}/2)$. $~p~$ denotes the four-momentum
\beq
  p_{\mu}(\hat{k}_{\mu}) = \frac{2}{a} \sin\left(\frac{\pi
      \hat{k}_{\mu}}{L}\right)
\label{eq:p-def}
\eeq
with the integer-valued lattice momentum
$~\hat{k}_{\mu} \in (-L/2, +L/2]$. $a$ is the lattice spacing.

The lattice ghost propagator is defined by inverting the Faddeev-Popov
(F-P) operator, the latter being the Hessian of the gauge functional
\Eq{eq:functional}. The F-P operator can be written in terms of the
(gauge-fixed) link variables $~U_{x,\mu}~$ as
\beq
  M^{ab}_{xy} = \sum_{\mu} A^{ab}_{x,\mu}\,\delta_{x,y}
  - B^{ab}_{x,\mu}\,\delta_{x+\hat{\mu},y}
  - C^{ab}_{x,\mu}\,\delta_{x-\hat{\mu},y}\quad
  \label{eq:FP-def}
\eeq
with
\beas
A^{ab}_{x,\mu} &=& \half~\delta_{ab}~\Tr\left[
                          U_{x,\mu}+U_{x-\hat{\mu},\mu} \right]\,,\\
B^{ab}_{x,\mu} &=& \half~ \Tr\left[ \sigma^b \sigma^a\, U_{x,\mu}\right]\,,
\\
C^{ab}_{x,\mu} &=& \half~ \Tr\left[ \sigma^a \sigma^b\, U_{x-\hat{\mu},\mu}
  \right]\,,
\eeas
where the $~\sigma^a\,, a=1,2,3~$ are the Pauli matrices.
In the continuum $~M^{ab}_{xy}~$ corresponds to the operator
$~M^{ab} = - \partial_{\mu} D^{ab}_{\mu}$, with $~D^{ab}~$
the covariant derivative in the adjoint representation.

The ghost propagator in momentum space is calculated from the
ensemble average
\bea
  G^{ab}(p) &=& \frac{1}{V} \sum_{x,y}
      \left\langle {\rm e}^{-2\pi i\,\hat{k}
      \cdot (x-y)} [M^{-1}]^{a\,b}_{x\,y}\right\rangle_U \\
            &=& \delta^{ab}~G(p)\,.
  \label{eq:G-def}
\eea
Following Ref.~\cite{Suman:1995zg, Cucchieri:1997dx} we have used the
conjugate gradient (CG) algorithm to invert $~M~$ on a plane wave
$~\vec{\psi}_c = \{~\delta_{ac} \exp (2\pi i\,\hat{k}\!\cdot\! x)~\}$.

After solving $~M \vec{\phi}=\vec{\psi}_c~$ the resulting
vector $~\vec{\phi}~$ is projected back on $~\vec{\psi}~$
so that the average $~G^{cc}(p)~$ over the color index $~c~$ can be
taken explicitly. Since the F-P operator $~M~$ is zero if acting on
constant modes, only $~\hat{k} \ne (0,0,0,0)~$ is permitted. Due to high
computational requirements to invert the F-P operator for each $~\hat{k}$,
separately, the estimators on a single, gauge-fixed configuration are
evaluated only for a preselected set of momenta $~\hat{k}$.

\section{Results}
\label{sec:results}

We consider various bare couplings in the interval
$~\beta=4/g_0^2~\in [2.1,2.5]~$ and lattice sizes up to $~20^4~$.
We compare the gluon and ghost propagators obtained with the
alternative gauge fixing methods SOR (`flips off') and FOR (`flips on')
both for the first (\fc{}) and best copy (\bc{}). In order to find the best
copies we always generate $~20~$ initial random gauge copies.

\begin{figure*}
\mbox{
\includegraphics[width=0.5\textwidth,height=0.48\textwidth]{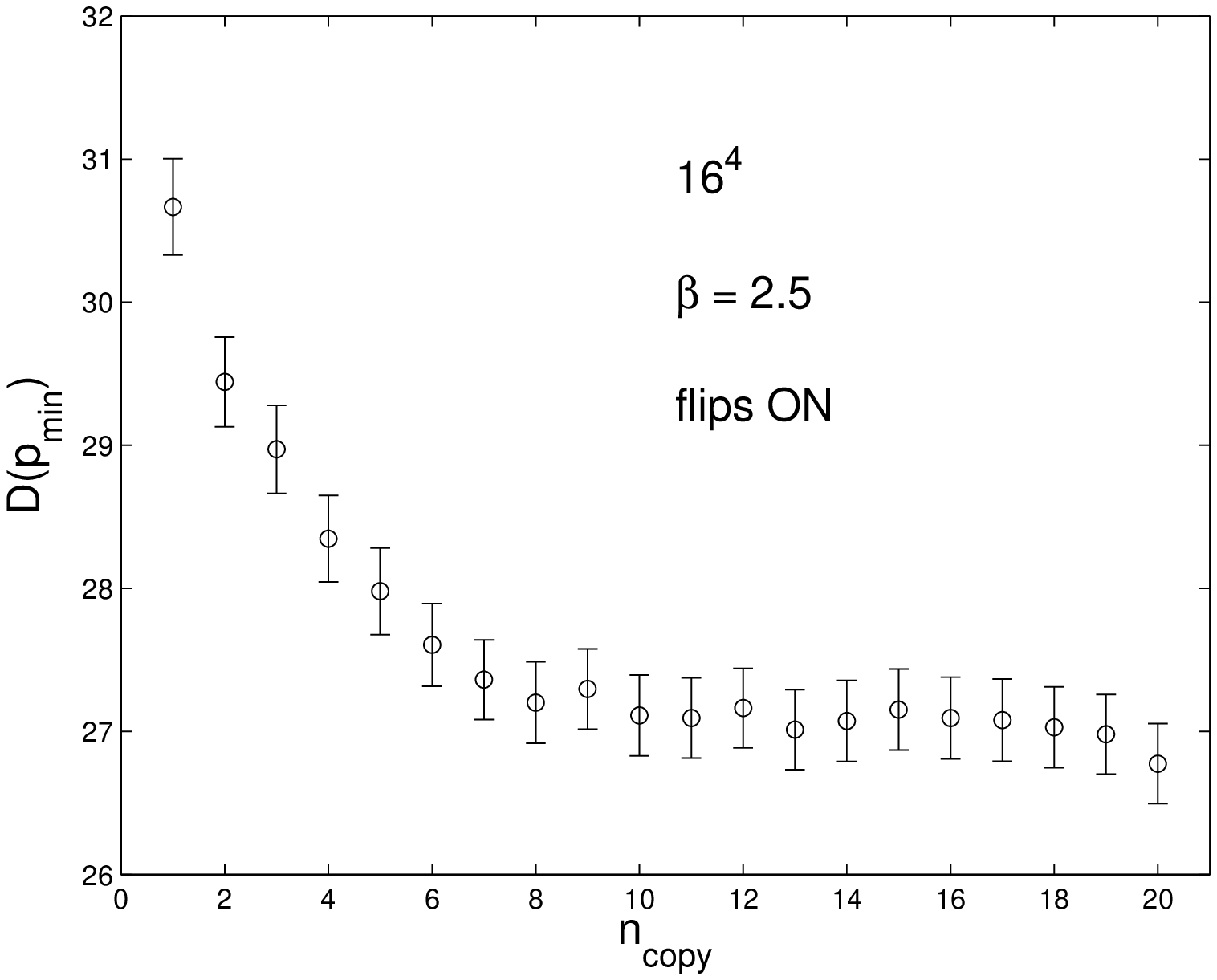}
\quad
\includegraphics[width=0.5\textwidth,height=0.48\textwidth]{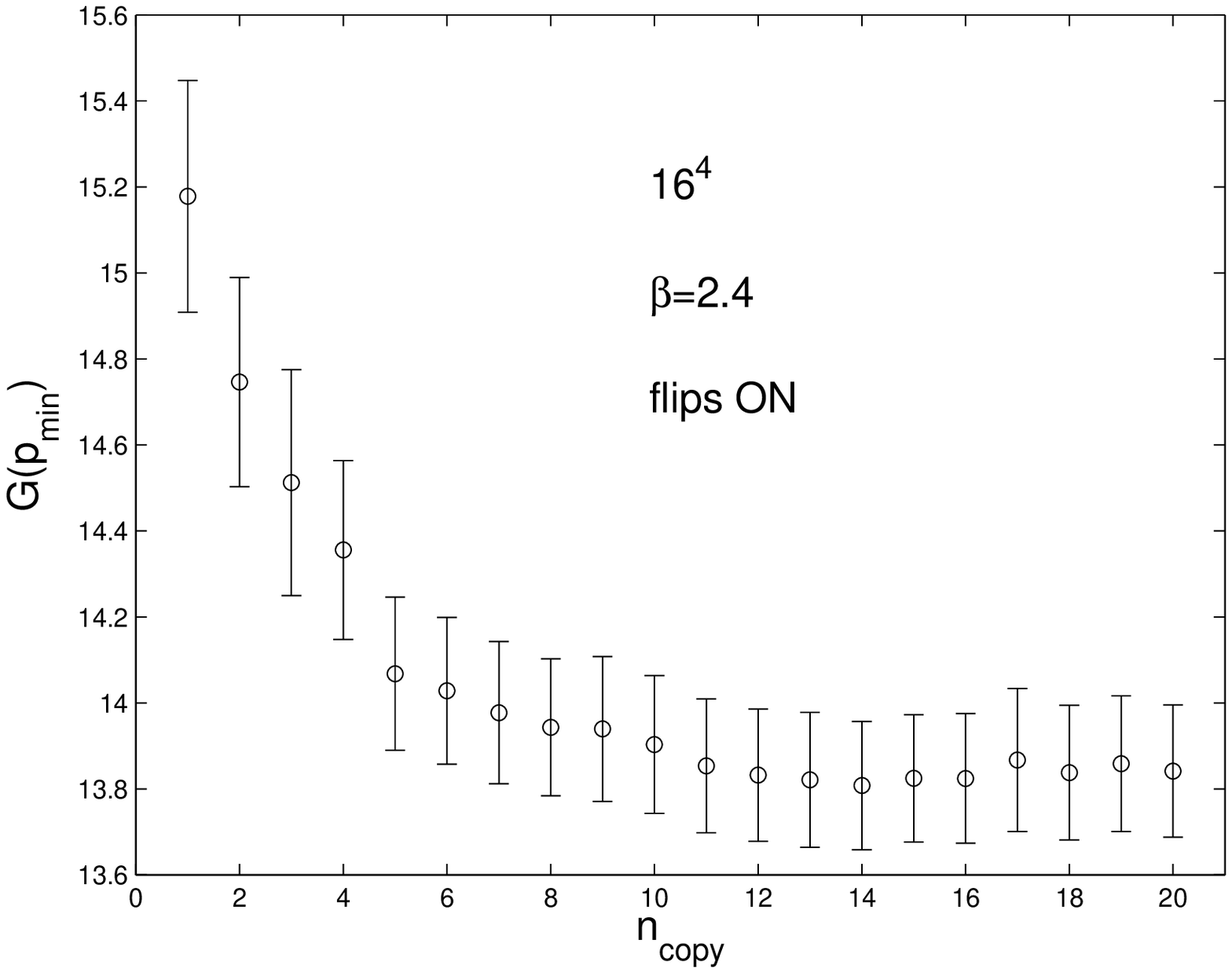}
}
\caption{
Gluon propagator (left) and ghost propagator (right) at lowest momentum
$~p_{min}=(2/a) \sin (\pi / L)~$
versus number of random copies employing the FOR method ('flips on') at
$~\beta=2.5~$ and $~2.4$, respectively (lattice size $~16^4~$).
}
\label{fig:Gl_Gh_copies}
\end{figure*}

In \Fig{fig:Gl_Gh_copies} we illustrate for the FOR method how fast
the gluon and ghost propagators are converging when determined from the
best copy out of the first $~n_{copy}~$ copies. We see plateaus occuring
for $~n_{copy} \ge O(10)$. We have convinced ourselves that
$~O(20)~$ copies are sufficient at least for $\beta \ge 2.3$ and lattice
sizes up to $20^4$.
For the SOR method the convergence is faster - although to worse
values of the gauge functional - such that
in principle a smaller number of copies would be sufficient within
the given parameter range.

Mostly we have concentrated on the lowest non-trivial on-axis lattice
momentum $~p_{min}~=~(2/a) \sin (\pi / L)~$ and some multiple on-axis momenta
in order to study the infrared limit for given lattice size and bare coupling.
We are aware of the fact that this choice is by far too restrictive in order
to get reliable results for the (renormalized) propagators in the continuum
and thermodynamic limit.

\begin{figure*}
\mbox{
\includegraphics[width=0.48\textwidth,height=0.46\textwidth]{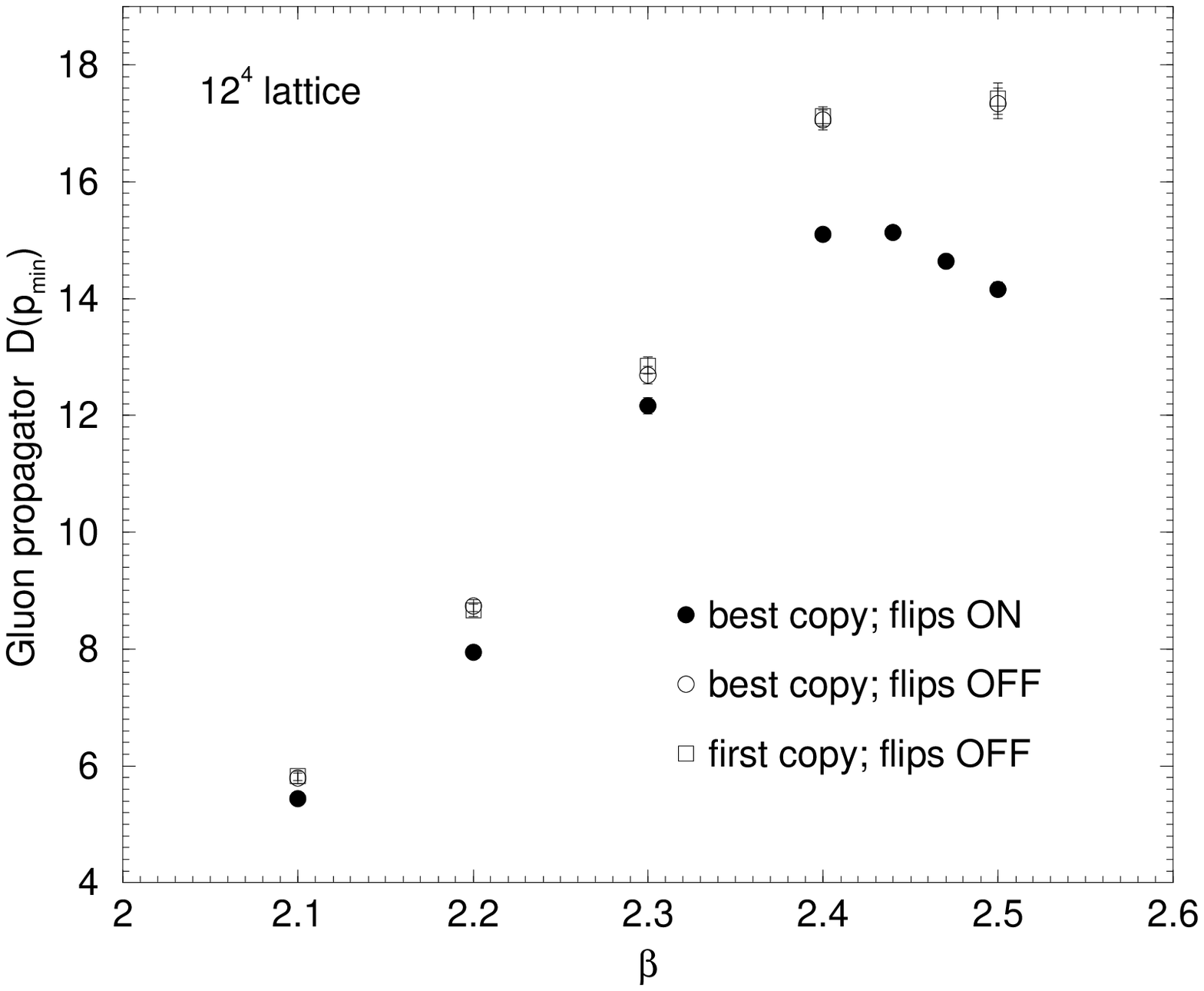}
\quad
\includegraphics[width=0.48\textwidth,height=0.47\textwidth]{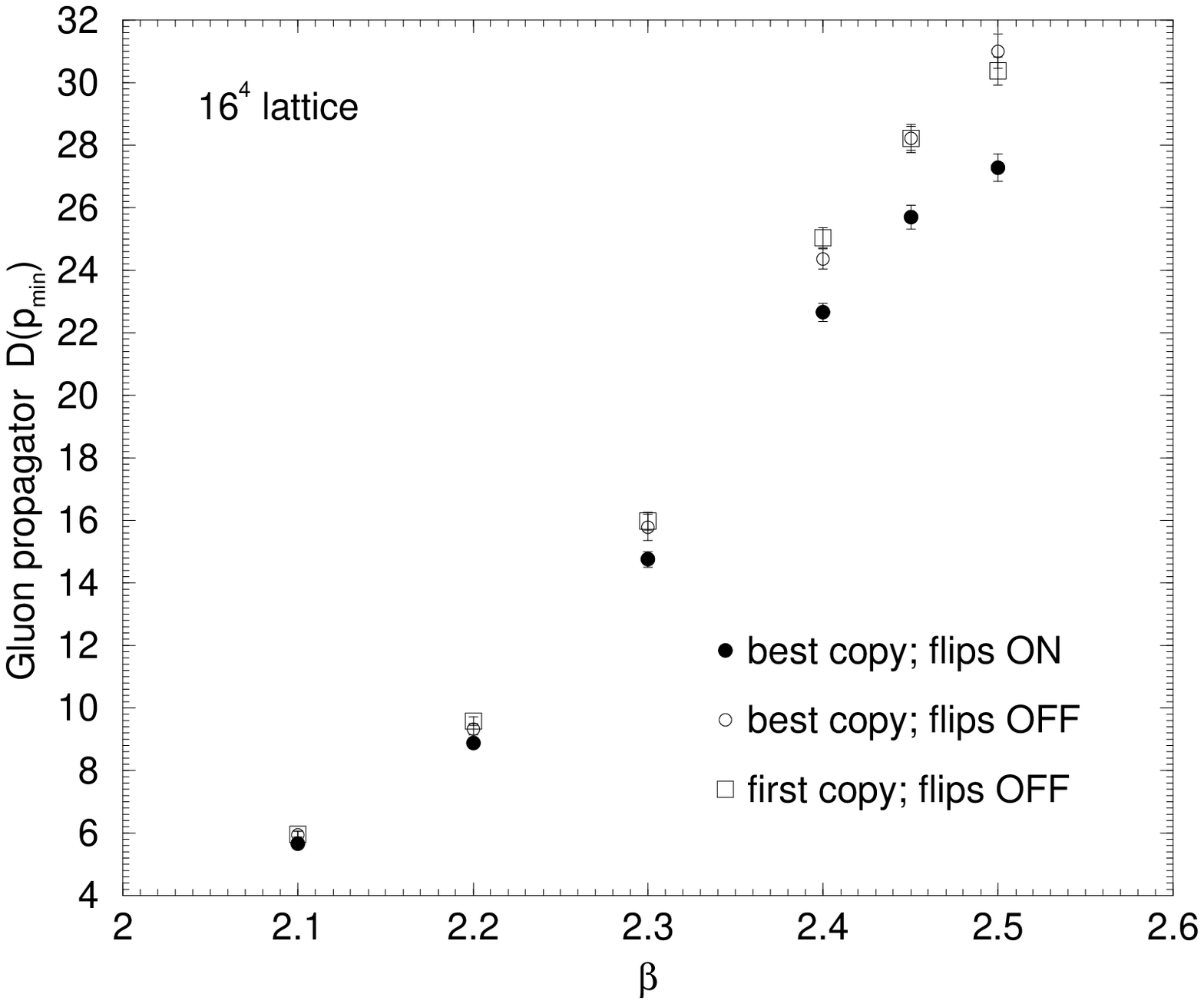}
}
\caption{
Gluon propagator $~D(p_{min})~$ at lowest momentum
for various $~\beta~$ and for lattice sizes $~12^4~$ (left) and $~16^4~$
(right). Full dots refer to FOR \fc{} and open squares (circles)
correspond to SOR \fc{} (\bc{}).
}
\label{fig:gluon_prop}
\end{figure*}

\begin{figure*}
\mbox{
\includegraphics[width=0.48\textwidth,height=0.48\textwidth]{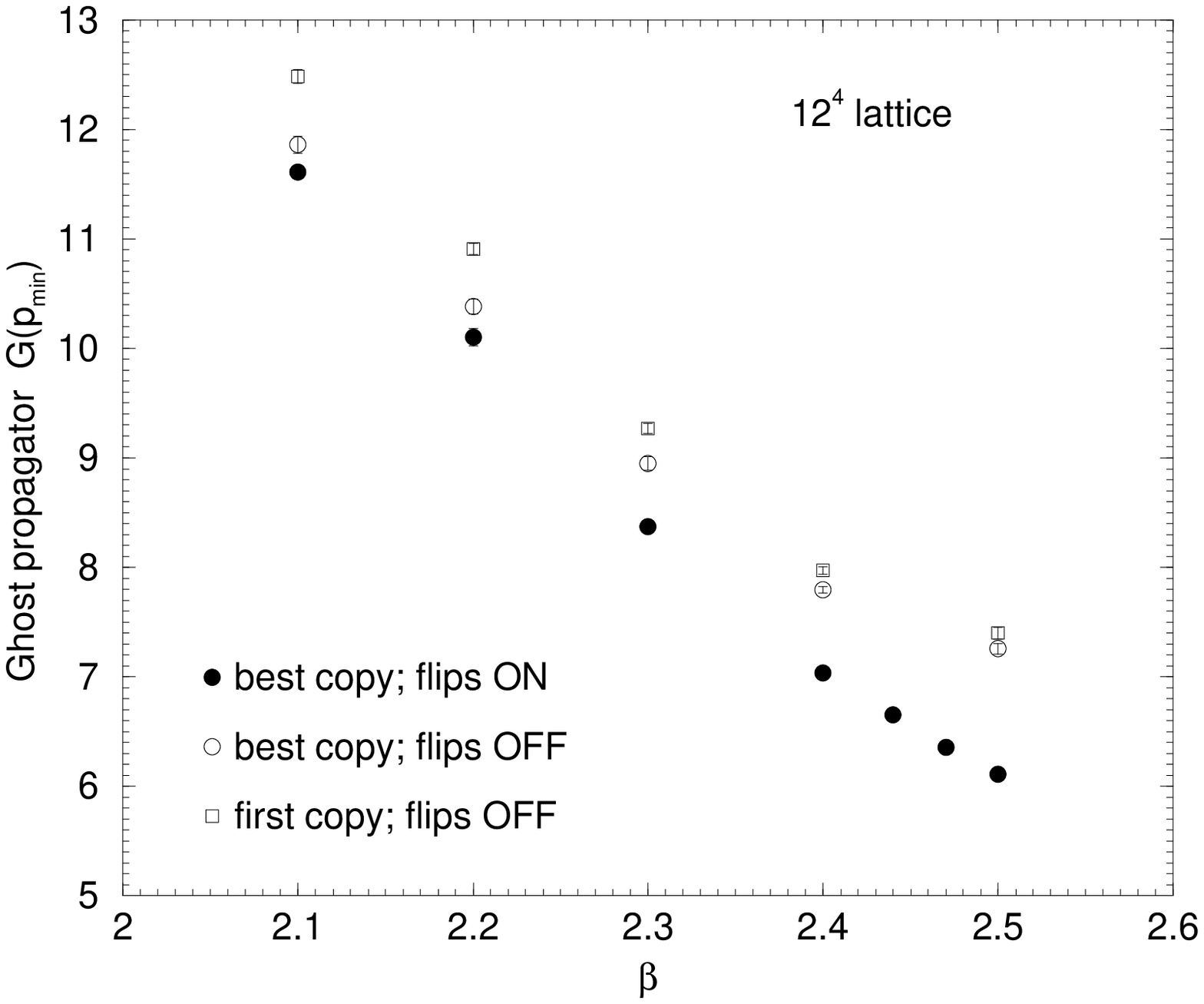}
\quad
\includegraphics[width=0.48\textwidth,height=0.48\textwidth]{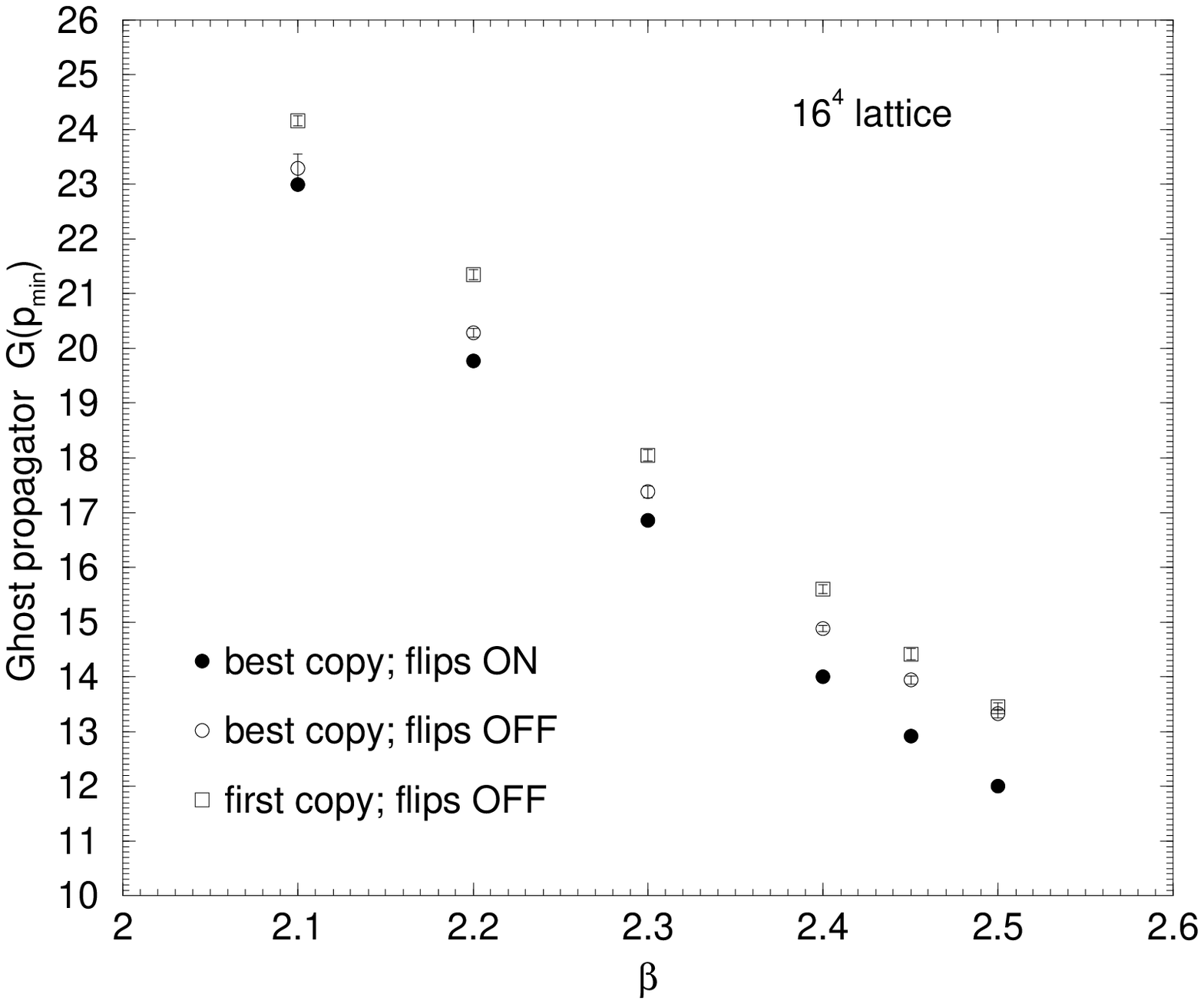}
}
\caption{
Ghost propagator $~G(p_{min})~$ as for \Fig{fig:gluon_prop}.
}
\label{fig:ghost_prop}
\end{figure*}

In \Fig{fig:gluon_prop} and \Fig{fig:ghost_prop} we show our results
for the lattice gluon $D(p_{min})$ and ghost propagators $G(p_{min})$ for
$12^4$ and $16^4$ lattices, always for the smallest non-vanishing momentum.
In order to demonstrate the effect of the $\mathbb{Z}(2)$ flips
in comparison with the SOR results obtained with \bc{} and \fc{} copies
\cite{Bakeev:2003rr} we show three sets of data points:
black dots correspond to FOR - `flips on' and \bc{} copies
and open circles (squares) correspond to  SOR - `flips off' for \bc{} (\fc{})
copies. The corresponding data are listed in \Tab{tab:Gl_Gh_prop}.

\begin{table*}
\begin{center}
\mbox{
\begin{tabular}{|c|c|c||c|c|c|}    \hline\hline
\multicolumn{6}{|c|}{$12^4$} \\ \hline\hline
$\beta$& $\#$ & $D_{\rm{FOR}}^{(bc)}$
       & $\#$ & $D_{\rm{SOR}}^{(bc)}$
                    & $D_{\rm{SOR}}^{(fc)}$
\\ \hline\hline
  2.10 & 1200 & $ 5.39  (6)$ &  900 & $ 5.79  (8)$ &  $ 5.83  (8)$
\\ \hline
  2.20 & 1200 & $ 7.94  (9)$ & 1200 & $ 8.74 (10)$ &  $ 8.66 (10)$
\\ \hline
  2.30 & 1200 & $12.16 (14)$ & 1200 & $12.69 (15)$ &  $12.85 (15)$
\\ \hline
  2.40 & 3600 & $15.10 (10)$ & 2080 & $17.06 (17)$ &  $17.12 (17)$
\\ \hline
  2.44 & 5100 & $15.13  (9)$ &      &                  &
\\ \hline
  2.47 & 5700 & $14.64  (9)$ &      &                  &
\\ \hline
  2.50 & 2650 & $14.16 (13)$ & 1760 & $17.34 (26)$ &  $17.42 (26)$
\\ \hline\hline
\multicolumn{6}{|c|}{$16^4$}\\    \hline\hline
$\beta$& $\#$ & $D_{\rm{FOR}}^{(bc)}$
       & $\#$ & $D_{\rm{SOR}}^{(bc)}$
                    & $D_{\rm{SOR}}^{(fc)}$
\\ \hline\hline
  2.10 & 1042 & $ 5.59  (7)$ & 918 & $ 5.93  (8)$ & $ 5.95  (8)$
\\ \hline
  2.20 &  900 & $ 9.01 (12)$ & 740 & $ 9.35 (14)$ & $ 9.58 (14)$
\\ \hline
  2.30 & 1100 & $14.88 (18)$ & 510 & $16.16 (31)$ & $15.97 (29)$
\\ \hline
  2.40 & 1032 & $22.65 (29)$ &1020 & $24.36 (32)$ & $25.03 (32)$
\\ \hline
  2.45 & 1020 & $25.69 (32)$ &1030 & $28.19 (36)$ & $28.21 (38)$
\\ \hline
  2.50 & 1040 & $26.86 (35)$ &1060 & $30.64 (44)$ & $30.37 (45)$
\\ \hline\hline
\end{tabular}
}
\hspace*{0.5cm}
\mbox{
\begin{tabular}{|c|c|c||c|c|c|}    \hline\hline
\multicolumn{6}{|c|}{$12^4$} \\ \hline\hline
$\beta$& $\#$ & $G_{\rm{FOR}}^{(bc)}$
       & $\#$ & $G_{\rm{SOR}}^{(bc)}$
                    & $G_{\rm{SOR}}^{(fc)}$
\\ \hline\hline
  2.10 & 1200 & $11.58 (4)$ & 900 & $11.87 (4)$ & $12.48 (7)$
\\ \hline
  2.20 & 1200 & $10.10 (8)$ &1200 & $10.39 (3)$ & $10.90 (5)$
\\ \hline
  2.30 & 1200 & $ 8.37 (2)$ &1200 & $ 8.99 (6)$ & $ 9.27 (4)$
\\ \hline
  2.40 & 3600 & $ 7.04 (1)$ &2080 & $ 7.80 (3)$ & $ 7.97 (4)$
\\ \hline
  2.44 & 5100 & $ 6.65 (1)$ &     &                  &
\\ \hline
  2.47 & 5700 & $ 6.36 (1)$ &     &                  &
\\ \hline
  2.50 & 2650 & $ 6.11 (1)$ &1760 & $ 7.26 (5)$ & $ 7.40 (5)$
\\ \hline\hline
\multicolumn{6}{|c|}{$16^4$}\\    \hline\hline
$\beta$& $\#$ & $G_{\rm{FOR}}^{(bc)}$
       & $\#$ & $G_{\rm{SOR}}^{(bc)}$
                    & $G_{\rm{SOR}}^{(fc)}$
\\ \hline\hline
  2.10  &1042 & $22.89 (6)$ & 918 & $23.12 (13)$ & $24.15 (8)$
\\ \hline
  2.20  & 900 & $19.83 (6)$ & 740 & $20.29  (6)$ & $21.34  (9)$
\\ \hline
  2.30  &1100 & $16.83 (5)$ & 510 & $17.27  (8)$ & $18.05 (10)$
\\ \hline
  2.40  &1032 & $14.00 (4)$ &1020 & $14.88  (6)$ & $15.60  (8)$
\\ \hline
  2.45  &1020 & $12.92 (5)$ &1030 & $13.86  (6)$ & $14.41 (11)$
\\ \hline
  2.50  &1040 & $12.02 (4)$ &1060 & $13.26  (6)$ & $13.45  (7)$
\\ \hline\hline
\end{tabular}
}
\end{center}
\caption{Data for the gluon propagator $D(p)$ (left) as well as for the
ghost propagator $G(p)$ (right) at lowest momentum $p=p_{min}$ obtained
with FOR (\bc{}) and SOR (\bc{} and \fc{}) methods on $12^4$  and $16^4$
lattices.
}
\label{tab:Gl_Gh_prop}
\end{table*}

We clearly see that the FOR method leads to an additional visible
Gribov copy effect not only for the ghost propagator but also for
the gluon propagator.
The effect is even more pronounced at higher $\beta$-values, i.e. at
smaller 'physical' lattice sizes.
We have convinced ourselves that this is compatible with the behavior of the
average maximal gauge functional $\langle F_{max} \rangle$. Its relative
difference determined with \bc{} copies for the FOR method
versus the SOR method is also rising with $\beta$. Later on
we shall see that this observation is also in one-to-one
correspondence with the gauge copy dependence for fixed $\beta$
and varying lattice size. The anatomy of the (new) FOR gauge
copies deserves further studies in the future. 

\begin{figure*}
\mbox{
\includegraphics[width=0.50\textwidth,height=0.48\textwidth]%
{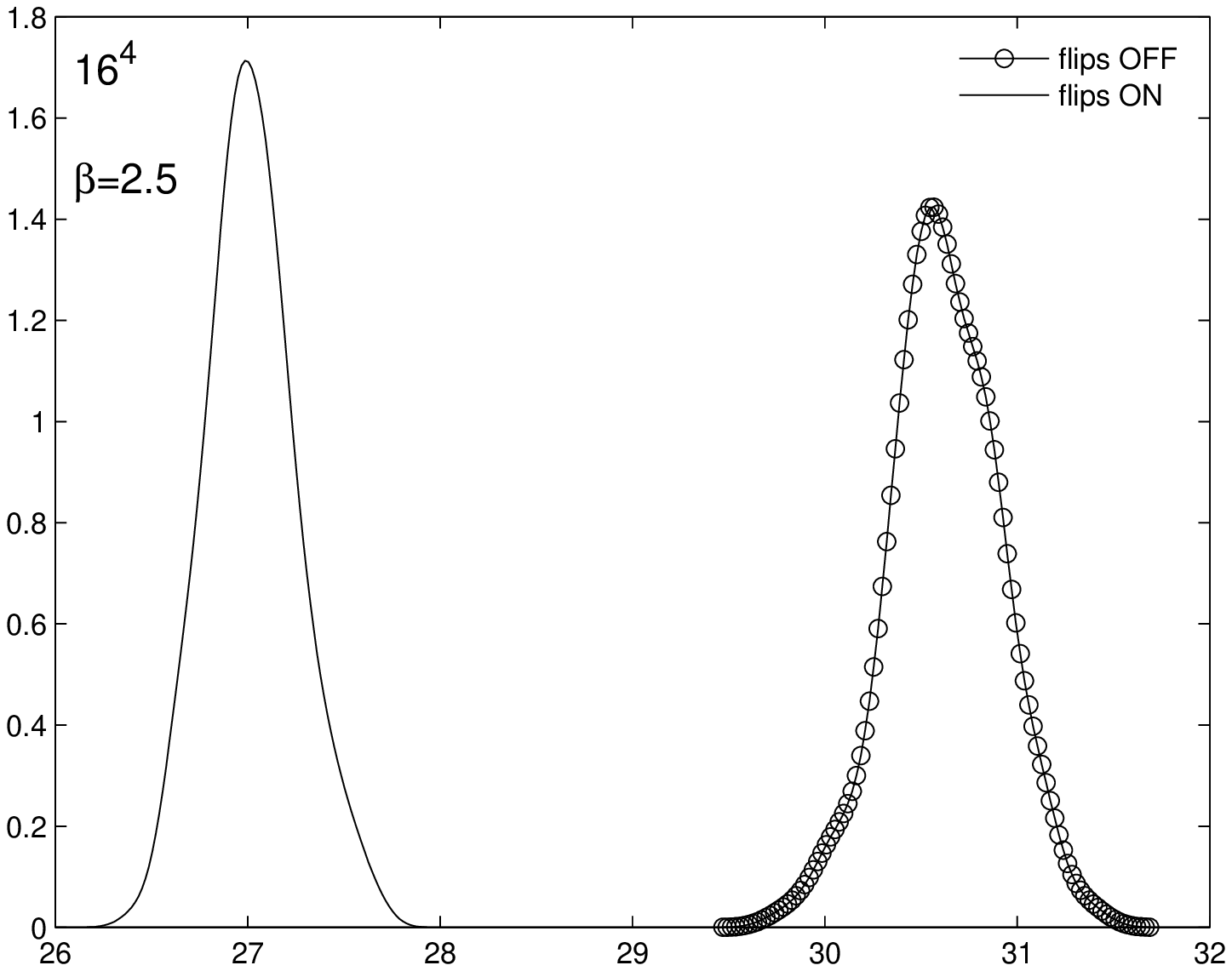}
\includegraphics[width=0.50\textwidth,height=0.48\textwidth]%
{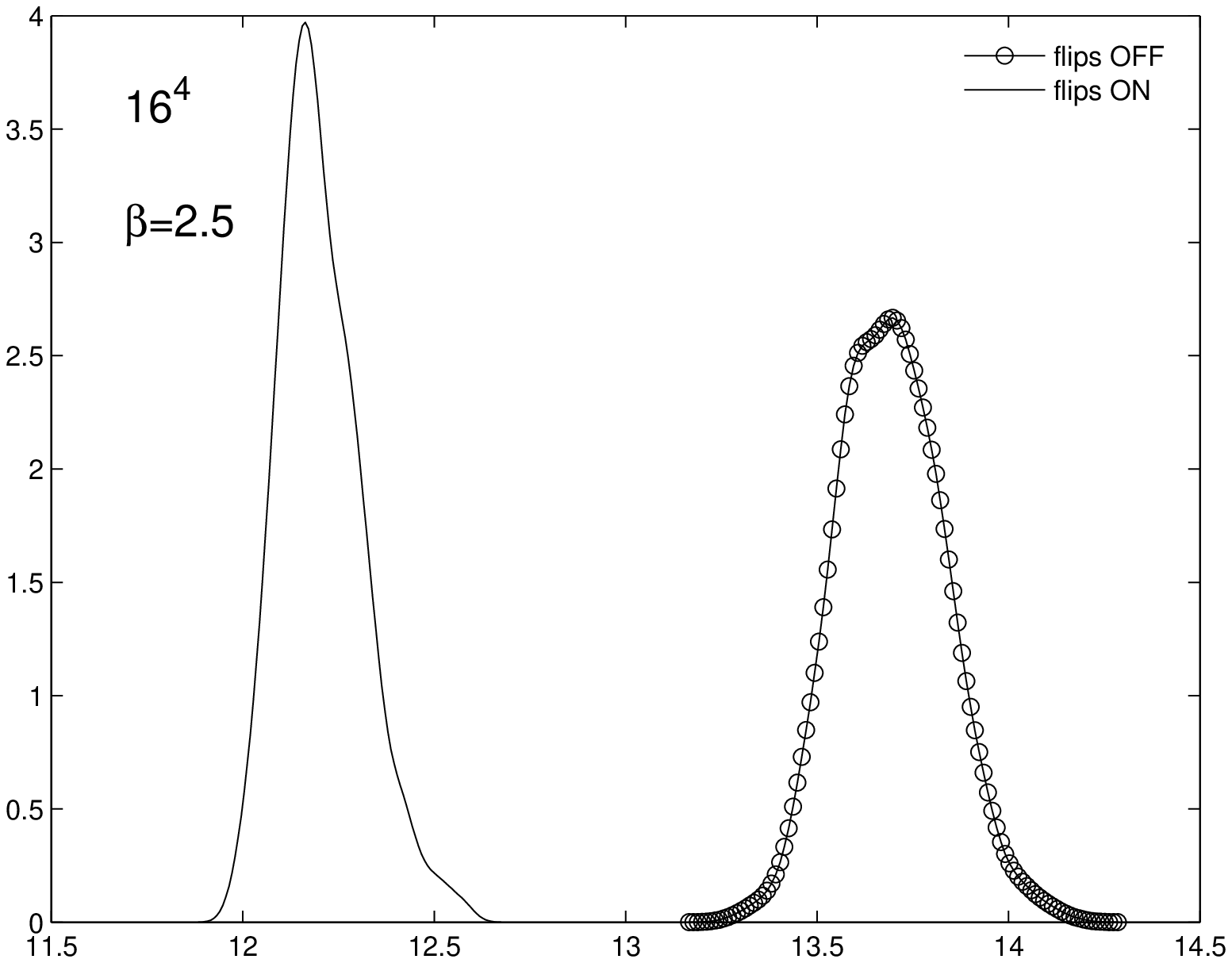}
}
\caption{SOR- and FOR-distributions for
$D^{(bc)}(p_{min})$ (left) and $G^{(bc)}(p_{min})$ (right)
at $\beta=2.5$ and $16^4$ lattice.}
\label{fig:Gl_Gh_dist}
\end{figure*}

In order to illustrate the strong Gribov copy effect in a slightly
different manner we compare smoothed distributions for the mean
value estimators for the gluon and ghost propagators for the \bc{} with
the FOR and SOR method, respectively (see \Fig{fig:Gl_Gh_dist}). The
mean value distributions have been obtained in accordance with the
bootstrap method \cite{efron} from replica of sequences of randomly
selected data.
Such bootstrapped resampling was applied to the initial MC data set as a
whole, the amount of replicas being typically 200. To smoothen the
distribution we have used the standard Nadaraya--Watson method with
normal kernel \cite{bowman}, and an improved Silverman's rule of thumb
for the choice of the corresponding bandwidth.

It is worth mentioning that the statistical errors for most of our data
have also been estimated through bootstrapped resampling.

\begin{figure*}
\mbox{
\includegraphics[width=0.50\textwidth,height=0.48\textwidth]{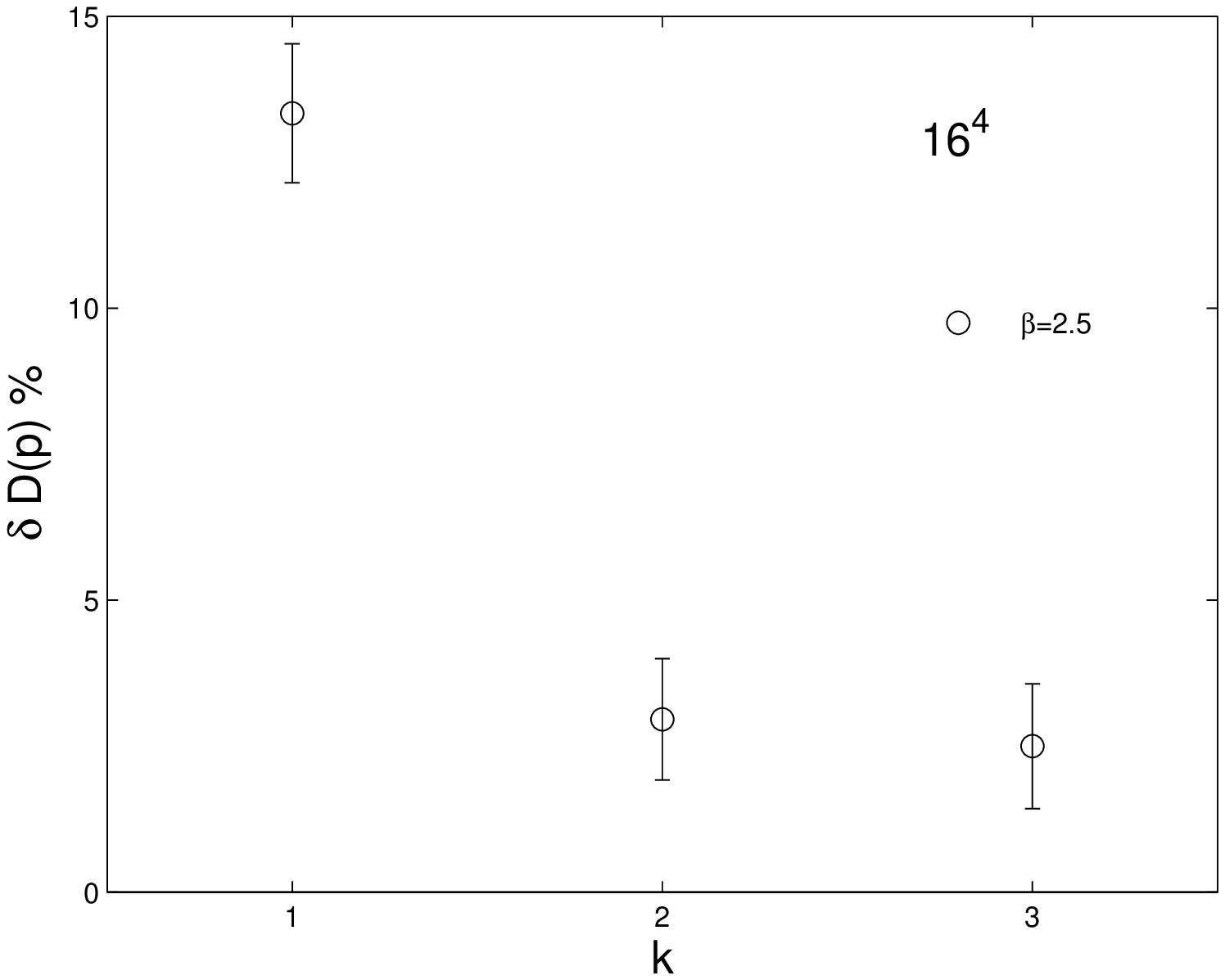}
\includegraphics[width=0.50\textwidth,height=0.48\textwidth]{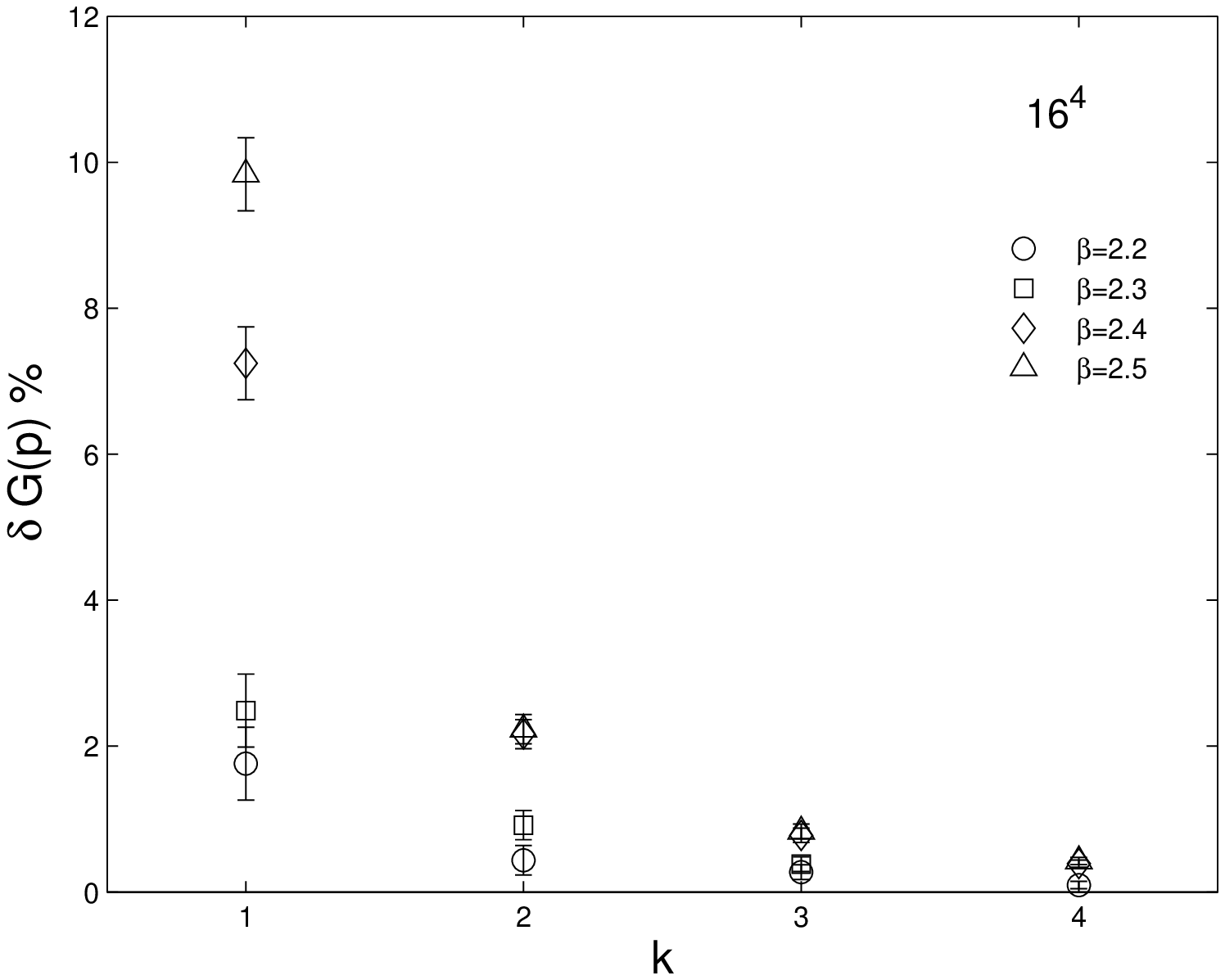}
}
\caption{
Left: relative deviation
$~\delta D(p) = (D_{\rm{SOR}}^{(bc)}-D_{\rm{FOR}}^{(bc)})/D_{\rm{FOR}}^{(bc)}~$
in percent for the gluon propagator at various (on-axis) lattice momenta
$~p(k)~$ (lattice size $16^4$, $\beta=2.5$). \\
Right: the analogous relative deviation for the ghost propagator for the
same lattice size but for $\beta=2.2, 2.3, 2.4$ and $2.5$.
}
\label{fig:Gl_Gh_rel_k}
\end{figure*}

We have also studied how the Gribov copy effect develops for larger
momenta $p(\hat{k})$. We have used multiples of the minimal lattice momentum
$\hat{k}=(0,0,0,k),~k=1,2,3,4~$ along one axis. We compare for the
gluon propagator the \bc{} SOR results with \bc{} FOR results in
terms of the relative deviation
\eq
\delta D(p)=(D_{\rm{SOR}}^{(bc)}-D_{\rm{FOR}}^{(bc)})/D_{\rm{FOR}}^{(bc)}~,
\en
\noi and analogously for the ghost propagator $~G(p)~$ at various $\beta$-values
and with fixed lattice size $~16^4~$ (see \Fig{fig:Gl_Gh_rel_k}).
For the gluon propagator our results are restricted to only one $\beta$-value
because of the much stronger statistical noise. Nevertheless,
the results presented for the gluon propagator point into the same direction
as for the ghost propagator. The effect of Gribov copies
still remains noticable at $p>p_{min}$, although decreasing for rising momenta.
The data for the ghost propagator at various momenta obtained from
independent Monte Carlo runs are also collected in \Tab{tab:Gh_p_16x16}.

\begin{table}[htb]
\begin{center}
\begin{tabular}{|c|c|c|c|c|c|c|}    \hline\hline
\multicolumn{7}{|c|}{FOR}   \\ \hline\hline
$\beta$& $\#$ & & $G(k=1)$ & $G(k=2)$ & $G(k=3)$ & $G(k=4)$ \\ \hline
 2.20 &400 & fc & 21.2(1) &  3.96(1)  & 1.510(2) & 0.8116(5)  \\ \hline
      &    & bc & 19.88(8)& 3.868(7)& 1.493(1) & 0.8076(4)  \\
\hline
 2.30 &400 & fc & 18.2(2) & 3.39(3)& 1.313(2)& 0.7276(6)   \\ \hline
      &    & bc & 16.88(8)& 3.267(6) & 1.299(1)& 0.7241(4)  \\
\hline
 2.40 &356 & fc & 15.4(1)& 2.87(1)&1.171(1) & 0.6693(3)  \\ \hline
      &    & bc & 13.8(1)&2.770(8) &1.156(2)  & 0.6647(4)  \\
\hline
 2.50 &400 & fc & 13.7(1)&  2.578(5)& 1.0897(8)& 0.6357(2) \\ \hline
      &    & bc & 12.2(1)& 2.508(5) & 1.079(1)& 0.6325(3) \\
\hline\hline
\multicolumn{7}{|c|}{SOR}   \\ \hline\hline
$\beta$& $\#$ &   & $G(k=1)$ & $G(k=2)$ & $G(k=3)$ & $G(k=4)$ \\ \hline
 2.20   & 200& fc &21.2(2)  &3.97(2)8 &1.511(3)  &0.8117(7) \\ \hline
        &    & bc &20.23(12)&3.885(10)&1.4971(25)&0.8084(7) \\
\hline
 2.30   & 200& fc &18.2(1)& 3.35(1)  &1.312(2) & 0.7272(5)  \\ \hline
        &    & bc &17.3(1)& 3.297(8) &1.304(1)& 0.7253(5)  \\
\hline
 2.40   & 370& fc & 15.6(1)& 2.87(1)& 1.171(1) & 0.6690(3) \\ \hline
        &    & bc & 14.8(1)& 2.83(1)& 1.165(1) & 0.6673(3) \\
\hline
 2.50   & 200& fc & 14.1(2)& 2.586(8) & 1.090(1)& 0.6359(4) \\ \hline
        &    & bc & 13.4(1)& 2.564(6) & 1.088(1)& 0.6352(3) \\
\hline
\hline
\end{tabular}
\end{center}
\caption{Ghost propagators $G(p)$ on the $16^4$ lattice for various
on-axis lattice momenta $~p(k)$.}
\label{tab:Gh_p_16x16}
\end{table}
We have also made a corresponding check for the gluon propagator at
zero momentum. On a lattice of size $20^4$ and for the same $\beta=2.5$ we
observed a deviation between the \bc{} FOR and SOR results of the order
$O(25 \%)$. This would of course have consequences for estimates like in
Ref.~\cite{Bonnet:2001uh,Boucaud:2006pc}, since the infinite volume
extrapolation of $D(0)$ there performed, although probably remaining finite,
will definitely suffer from uncontrolled systematic uncertainties.

It is interesting to study the volume dependence of the Gribov
copy effect, in view of Zwanziger's recent claim mentioned at the beginning
\cite{Zwanziger:2003cf}.
\begin{figure}[htm]
\mbox{
\includegraphics[width=0.47\textwidth,height=0.47\textwidth]%
{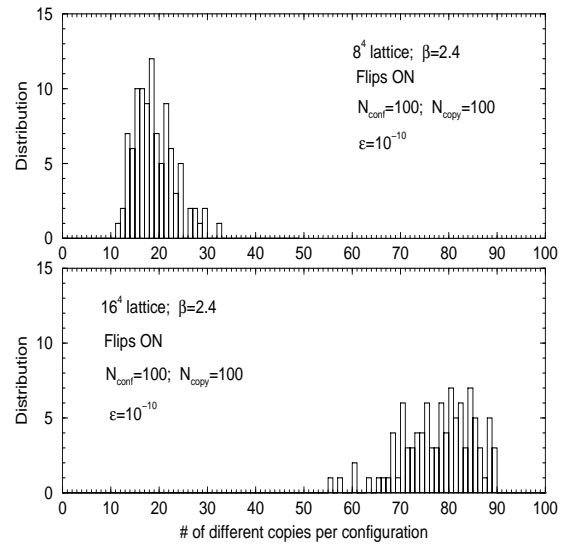}
}
\caption{Distributions of the number of different gauge copies found
with the FOR method at $\beta=2.40$ for lattice sizes $8^4$ and $16^4$.
}
\label{fig:no_copies}
\end{figure}
\begin{figure}[htm]
\mbox{
\includegraphics[width=0.47\textwidth,height=0.47\textwidth]%
{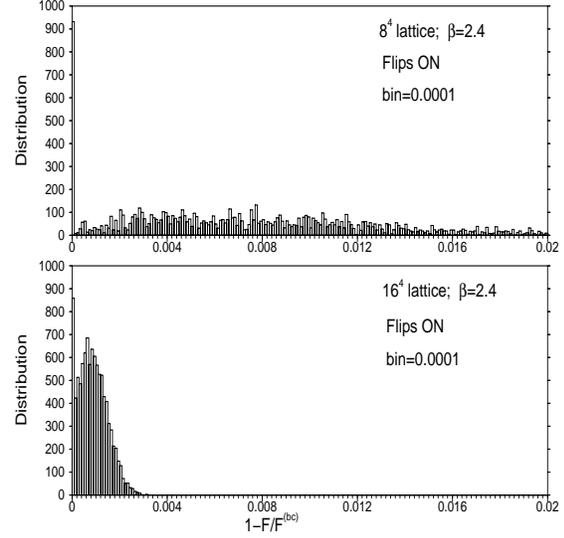}
}
\caption{Distributions of the deviation of the gauge functional values
for different gauge copies relative to the best copy per configuration.
FOR method at $\beta=2.40$ for lattice sizes $8^4$ and
$16^4$.
}
\label{fig:func_copies}
\end{figure}
\begin{figure}[htm]
\mbox{
\includegraphics[width=0.47\textwidth,height=0.47\textwidth]%
{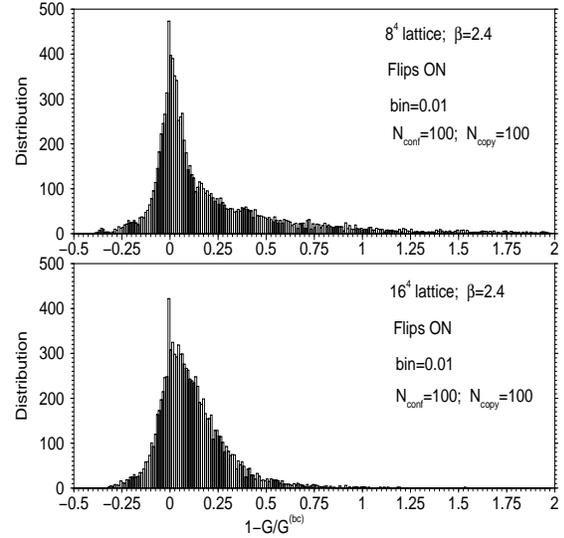}
}
\caption{Distributions of ghost propagator values
at lowest non-trivial momentum for different gauge copies as in
\Fig{fig:func_copies}.
}
\label{fig:ghost_copies}
\end{figure}
\begin{figure}[htm]
\mbox{
\includegraphics[width=0.50\textwidth,height=0.42\textwidth]%
{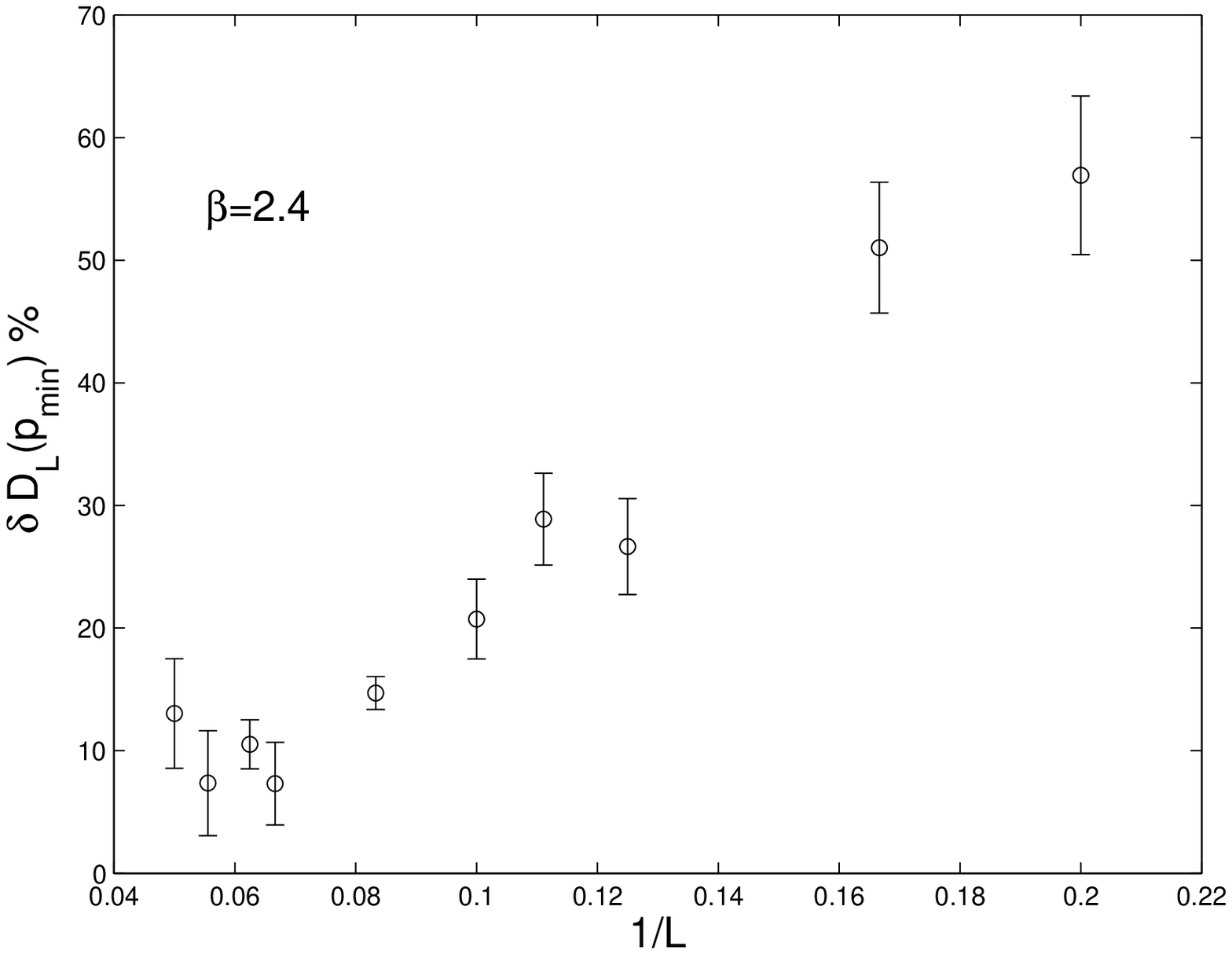}
}
\caption{
Relative deviation  $~\delta D_L(p_{min})~\equiv
~(D_{\rm{SOR}}^{(fc)}-D_{\rm{FOR}}^{(bc)})/D_{\rm{FOR}}^{(bc)}~$
in percent for the gluon propagator $~D~$ for various linear lattice
sizes $~L~$ and smallest non-vanishing momentum
$~p_{min}~=~(2/a) \sin(\pi/L)~$ ($\beta=2.4$).
}
\label{fig:Gl_rel_L}
\end{figure}
First of all we have convinced ourselves that the number of gauge copies
is strongly rising with the lattice volume as it should be. This is
clearly demonstrated in \Fig{fig:no_copies} providing the distributions
of the number of gauge copies per configuration found with the FOR method
(`flips on') for lattice sizes $8^4$ and $16^4$ at $\beta=2.40$.
In both cases we have generated 100 configurations with 100 gauge copies
each. It turns out that identical (or degenerated) copies can be well
recognized at an accuracy for the gauge functional \Eq{eq:functional}
of $O(10^{-10})$. Adjacent copies normally differ in the values for the
gauge functional at a level of $O(10^{-6})$. Now let us compare the
distributions of the corresponding values of the functional $F$ for each copy
found. In order to normalize the values  with respect to the highest
(i.e. best) value per configuration we show the relative deviation
$(F_{max}^{(bc)}-F_{max})/F_{max}^{(bc)}$. The frequency distributions of
these values are shown in \Fig{fig:func_copies} for the same ensembles as
used for  \Fig{fig:no_copies}.
There is a very clear tendency that the variance of
the gauge functional becomes much smaller if we increase the lattice volume.
A similar tendency becomes visible in \Fig{fig:ghost_copies}, where we plot
for the same set of configurations and gauge copies the distributions
for the single values of the ghost propagator for the lowest non-vanishing
on-axis momentum. Also in this case we have normalized the single values
as  $(G^{(bc)}-G)/G^{(bc)}$, i.e.
taking the relative deviation of the propagator at a given copy $G$ from
the value computed on the best copy $G^{bc}$, the latter chosen again with
respect to the gauge functional value. We see that the long tail seen for
the smaller lattice disappears for the larger lattice.
Although the fact that close values of the gauge functional will not tell
anything about how much the corresponding gauge configurations are differing
from each other (irrespective of a global relative gauge transformation)
we would like to interprete our finding of shrinking distributions as a
weakening of the Gribov problem with increasing 'physical' lattice size.

\begin{figure*}
\mbox{
\includegraphics[width=0.50\textwidth,height=0.45\textwidth]%
{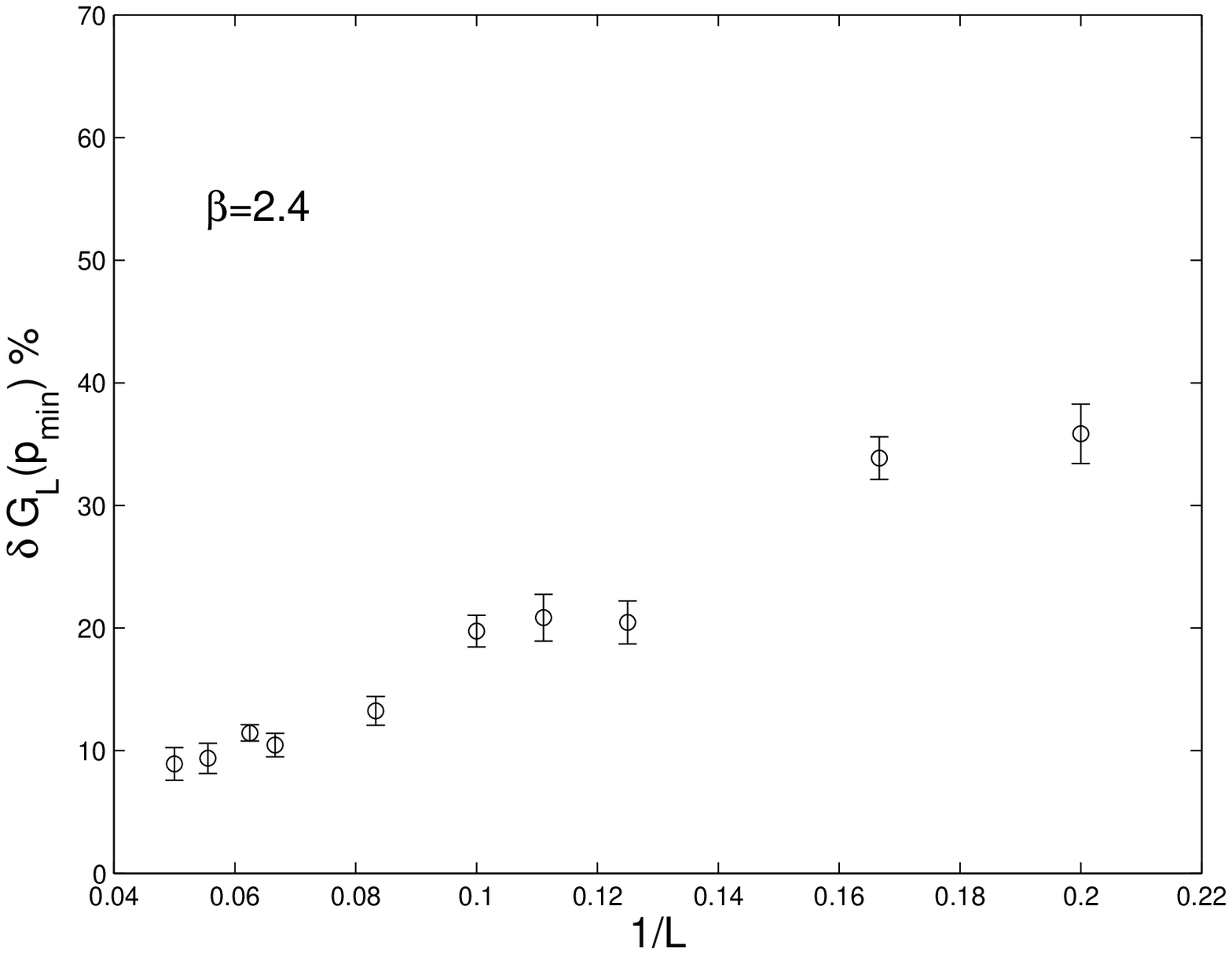}
\includegraphics[width=0.50\textwidth,height=0.45\textwidth]%
{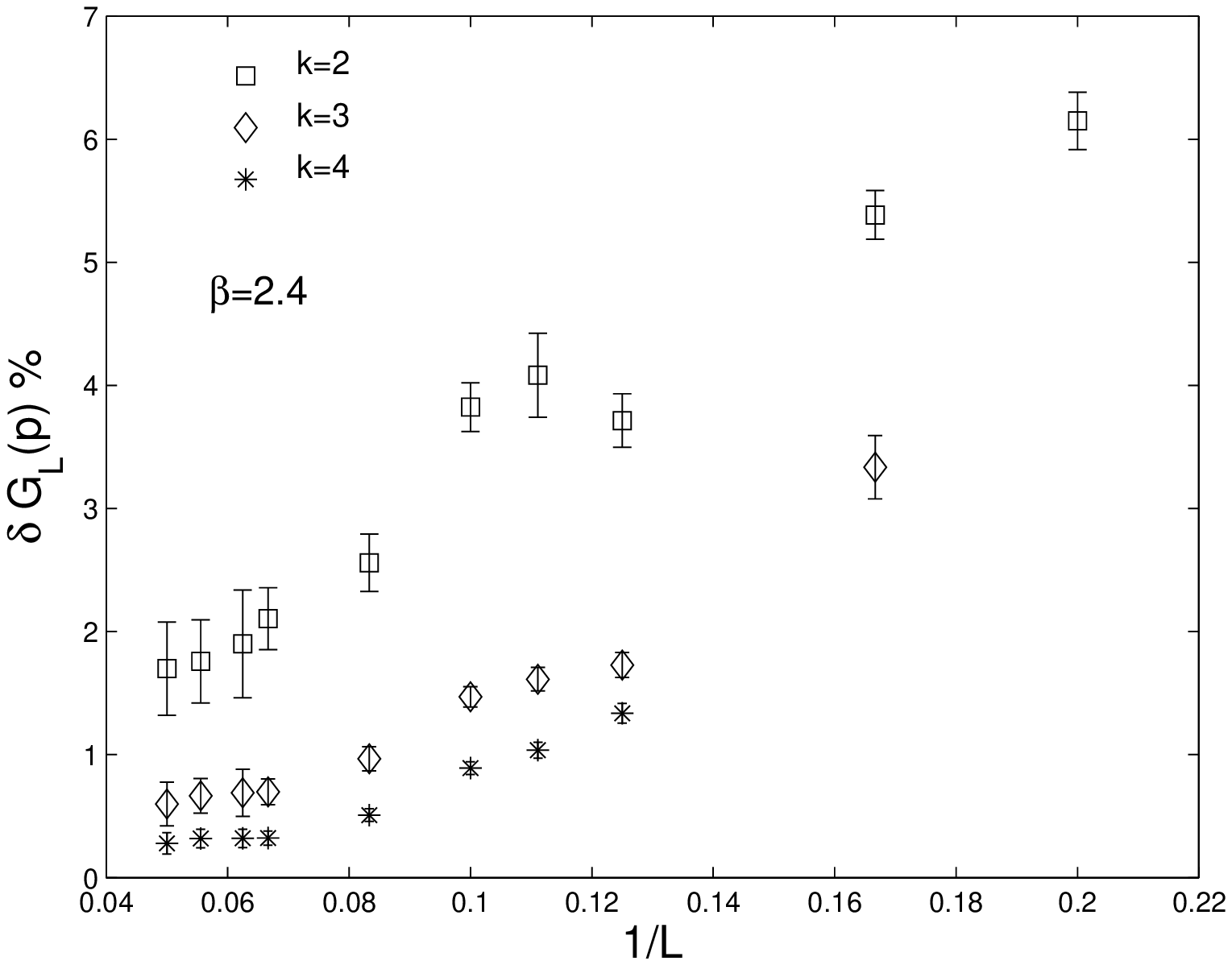}
}
\caption{
Relative deviation  $~\delta G_L(p)~\equiv
~(G_{\rm{SOR}}^{(fc)}-G_{\rm{FOR}}^{(bc)})/G_{\rm{FOR}}^{(bc)}~$
in percent for the ghost propagator $~G~$ at $\beta=2.4$
for various linear lattice sizes $~L~$ and the smallest non-vanishing
momentum $~p_{min}~=~(2/a) \sin(\pi/L)~$ (left) as well as for
on-axis momenta \mbox{$~p(k), ~k=2,3,4$ (right)}.
}
\label{fig:Gh_rel_Lk}
\end{figure*}

Moreover, we have plotted the relative deviation
\eq
\delta D_L(p_{min})~\equiv
(D_{\rm{SOR}}^{(fc)}-D_{\rm{FOR}}^{(bc)})/D_{\rm{FOR}}^{(bc)}
\en
for the gluon propagator (see \Fig{fig:Gl_rel_L}) and analogously for the
ghost propagator (see l.h.s. of \Fig{fig:Gh_rel_Lk})
as a function of the inverse linear lattice size $~1/L~$,
both determined at the minimal momentum $~p_{min}$. Here we have
used data for fixed $~\beta=2.4~$ and lattice sizes from $~L=5~$ up
to $~L=20$.
In close correspondence to our observations presented in Figs.
\ref{fig:gluon_prop} and \ref{fig:ghost_prop}
we see that the Gribov copy effect becomes weaker (stronger)
for increasing (decreasing) 'physical' lattice size and
correspondingly decreasing (increasing) minimal momentum,
at least up to a certain value of the lattice size ($\aleq 15 $).
One would of course need larger values of $L$ to make a reliable
conclusion about the limit $L\to\infty$. Anyway, at our largest
lattice value $L=20$ the Gribov copy effect is still quite strong.

For the ghost propagator, where the signal to noise ratio is more favourable,
we have found an analogous behavior also for the multiple on-axis momenta
$~k=2,3,4$ (see r.h.s. of \Fig{fig:Gh_rel_Lk}).

\begin{figure*}
\mbox{
\includegraphics[width=0.50\textwidth,height=0.50\textwidth]%
{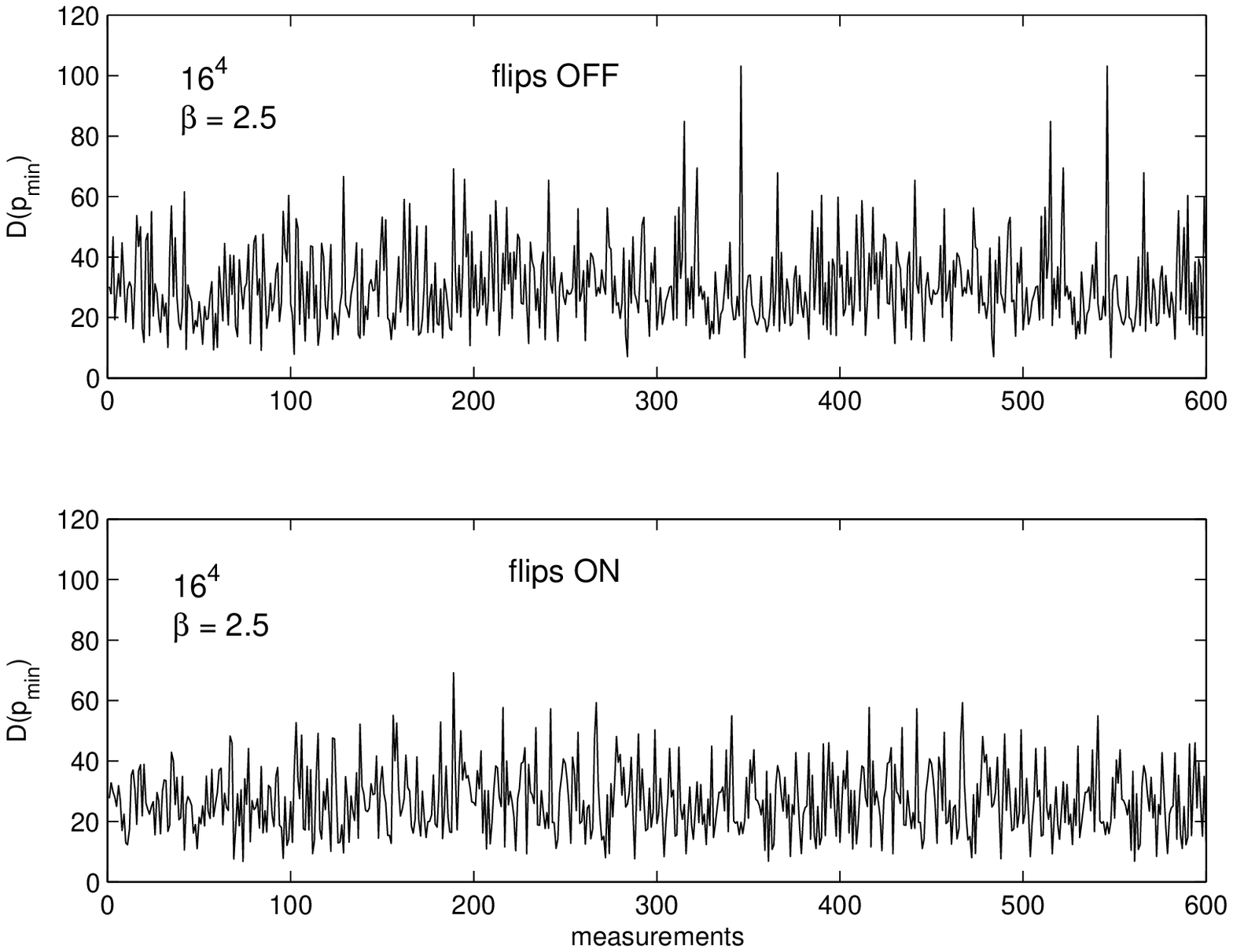}
\includegraphics[width=0.50\textwidth,height=0.50\textwidth]
{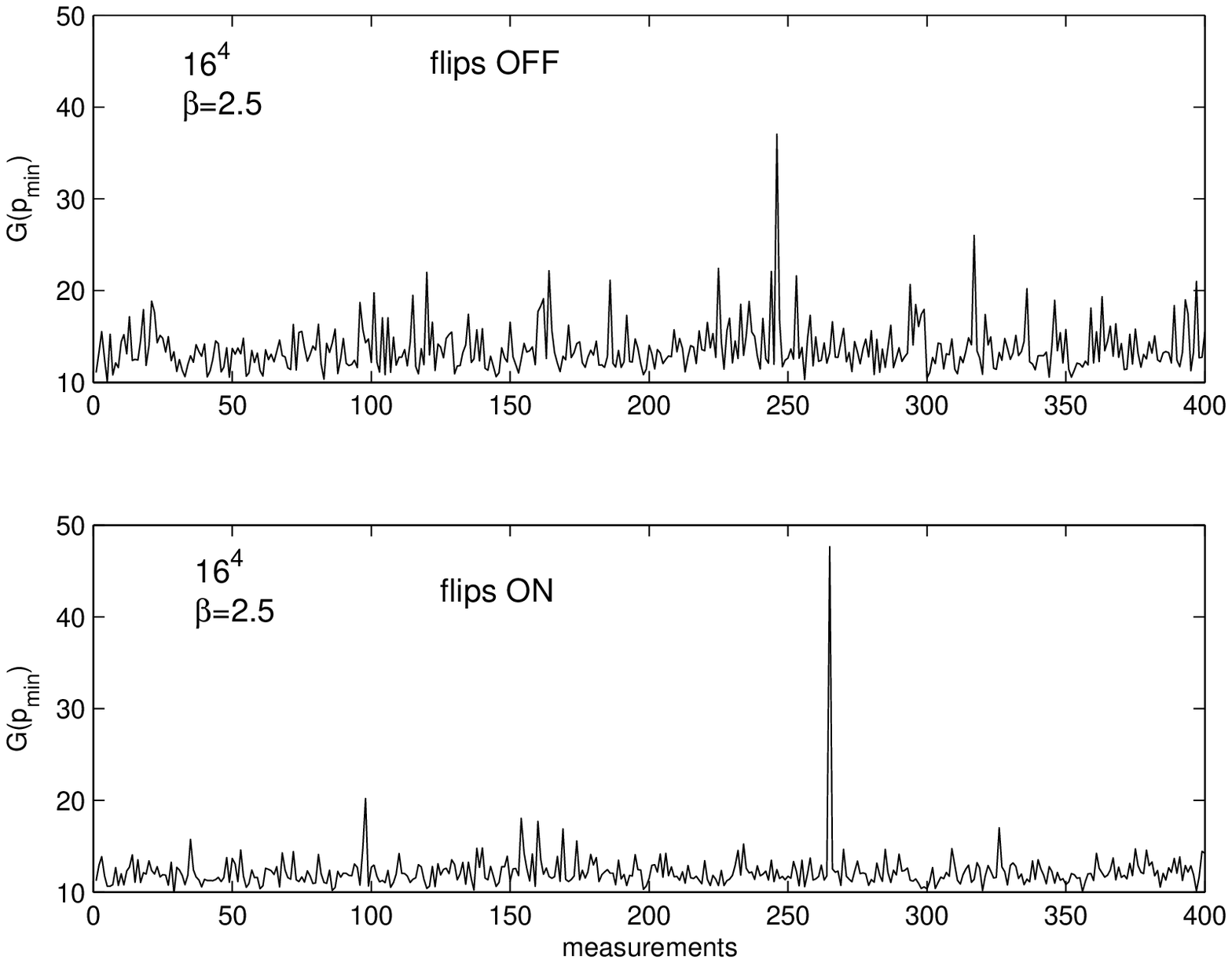}
}
\caption{
Time histories for $D^{(bc)}(p_{min})$ (left) and $G^{(bc)}(p_{min})$
(right) for both SOR and FOR methods at $\beta=2.5$ and $16^4$ lattice.}
\label{fig:Gl_Gh_hist_16x16_b2p50}
\end{figure*}

In \cite{Bakeev:2003rr} two of us have reported on rare Monte Carlo events
with exceptionally large values of the ghost propagator occuring for the
SOR gauge fixing method for larger $~\beta$ values. In
\Fig{fig:Gl_Gh_hist_16x16_b2p50} we show some time histories for the
gluon and ghost propagators for $\beta=2.5$ and a $16^4$ lattice, comparing
\bc{} SOR with \bc{} FOR. We see that for the `best copy - flips on' case (FOR) the fluctuations for both propagators are smaller. But for the ghost
propagator the effect of exceptionally large values, in general related to
small eigenvalues of the F-P operator \cite{Sternbeck:2005vs}, is still there.

\begin{figure*}
\mbox{
\includegraphics[width=0.50\textwidth,height=0.50\textwidth]%
{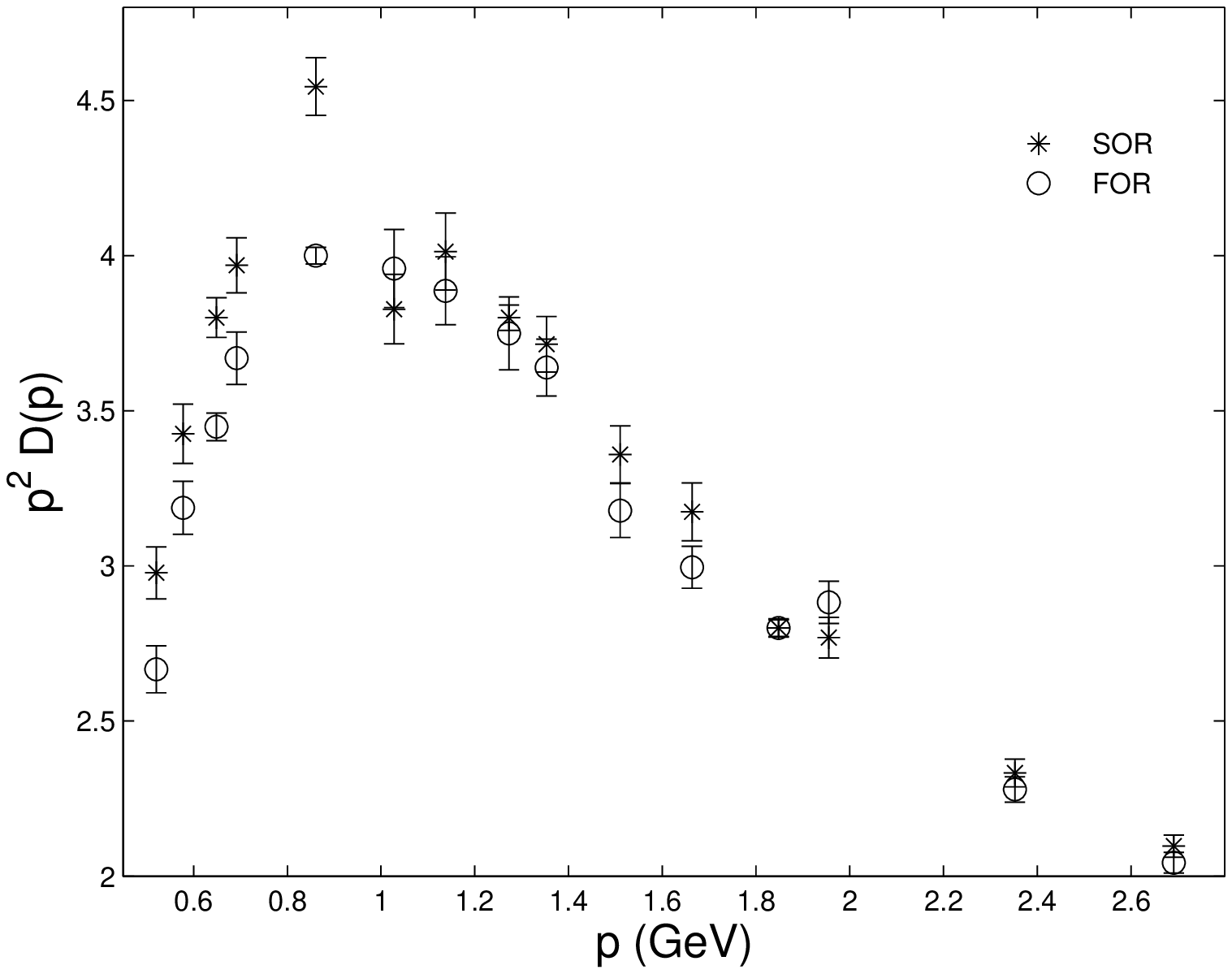}
\includegraphics[width=0.50\textwidth,height=0.50\textwidth]
{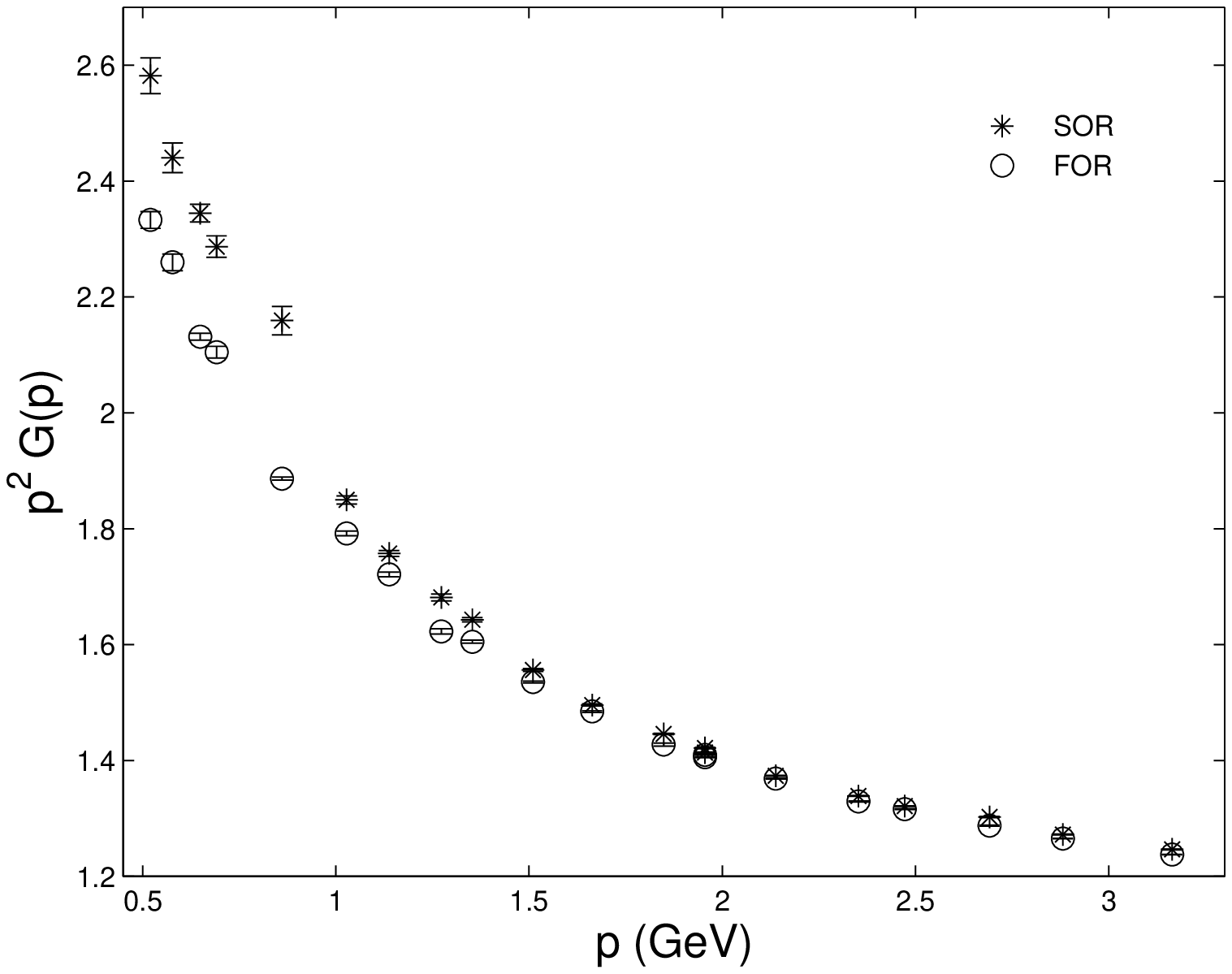}
}
\caption{
Gluon form factor $~p^2 D(p)~$ (left) and ghost form factor $~p^2 G(p)~$
(right) both for the \bc{} SOR and \bc{} FOR methods versus momentum
obtained for various lattice sizes and fixed $~\beta=2.4$.
}
\label{fig:Gl_Gh_scale}
\end{figure*}

Concluding we show the form factors of the gluon propagator
$~p^2 D(p)~$ and of the ghost propagator $~p^2 G(p)~$ in physical units
as a function of the physical momentum for fixed $~\beta=2.4~$ and lattice
sizes varying from $~10^4~$ to $~20^4~$. We have rescaled the gluon propagator
values $~D(p)~$ with factors $~a^2~$ and $~g_0^2~$
and the ghost propagator $~G(p)~$ with $~a^2~$, respectively, in order to
translate to the corresponding continuum (bare) propagators (compare with
\cite{Bloch:2003sk}). To estimate the lattice spacing in physical units
we have used the string tension: $~a^2 \sigma = .071~$
\cite{Fingberg:1992ju} with the standard value
$~\sqrt{\sigma} = 440 \mathrm{MeV}$. The form factor results for both
methods \bc{} SOR and \bc{} FOR are shown together in \Fig{fig:Gl_Gh_scale}.
Again the figure shows clear Gribov copy effects for both the propagators
and not only for the ghost propagator. We did not apply any
overall renormalization here. The statistics
collected for these runs is listed in \Tab{tab:stat}.
\begin{table}[htb]
\begin{center}
\begin{tabular}{|c|c|c|c|c|c|c|c|c|c|c|}    \hline\hline
\multicolumn{11}{|c|}{FOR}   \\ \hline\hline
$L$&5 &6 &8  & 9 & 10 &12&15&16&18&20\\ \hline
$N_{conf}$ &1000 & 1000 & 800 & 600  & 600 & 500&400&356&200&200  \\
\hline\hline
\multicolumn{11}{|c|}{SOR}   \\ \hline\hline
$L$&5 &6 &8  & 9 & 10 &12&15&16&18&20\\ \hline
$N_{conf}$ &1000 & 1000 & 800 & 500  & 500 & 400&400&370&100&100  \\
\hline
\hline
\end{tabular}
\end{center}
\caption{Statistics for the measurements at different $L$ and $\beta=2.4$.}
\label{tab:stat}
\end{table}

\section{Conclusions}
\label{sec:conclusions}

In this paper we have demonstrated that there is a visible Gribov problem
for the ghost propagator as well as for the gluon propagator computed
in $SU(2)$ lattice gauge theory within the Landau gauge. In order to show
this we have enlarged the gauge orbits of given Monte Carlo generated
gauge fields by non-periodic $\mathbb{Z}(2)$ transformations,
flipping all links in a given direction on a slice orthogonal to that.
This allows a preconditioning which maximizes the gauge functional before
applying the overrelaxation algorithm.

We have found indications for a weakening of the Gribov copy effect
both going to larger momenta at fixed volume and also increasing the
lattice size $L$ while correspondingly lowering the minimal non-zero
momentum, at least up to a certain value of the lattice size ($\aleq 15$).
However, one would need larger values of $L$ to draw a reliable
conclusion about the  limit $L\to\infty$.

We have not shown the momentum scheme running coupling which can
be determined from the form factors of the propagators discussed
here assuming that the renormalization factor for the ghost-gluon
vertex is constant. This will be discussed in a future paper,
where we want to present data for larger lattices and a larger
spectrum of (off-axis) momenta.

\section*{ACKNOWLEDGEMENTS}
This investigation has been supported by the Heisenberg-Landau program
of collaboration between the Bogoliubov Lab of Theoretical Physics of the
Joint Institute for Nuclear Research Dubna, Russia and german institutes.
V.K.M. acknowledges support by an RFBR grant 05-02-16306.
G.B. acknowledges support from an INFN fellowship. M.M.-P. thanks the DFG
for support under grant FOR 465 / Mu932/2-2.

\bibliographystyle{apsrev}

\end{document}